\documentclass[letterpaper, 10 pt, conference]{ieeeconf}  % Comment this line out if you need a4paper

\IEEEoverridecommandlockouts                              % This command is only needed if 
\usepackage{graphics} % for pdf, bitmapped graphics files
\usepackage{times} % assumes new font selection scheme installed
\usepackage{amsmath} % assumes amsmath package installed
\usepackage{amssymb}  % assumes amsmath package installed
\usepackage{booktabs}
\usepackage{url}
\usepackage{hyperref}
\usepackage{xcolor}
\usepackage{float}
\usepackage{optidef}
\usepackage{cite}
\usepackage{mathtools}
\usepackage{nomencl}
\usepackage{array}
\usepackage{caption}
\usepackage{subcaption}

\usepackage{amsthm}
\theoremstyle{remark}
\newtheorem{remark}{Remark}
\usepackage{multirow}
\usepackage[ruled,vlined]{algorithm2e}
\usepackage{optidef}
\usepackage{hhline}
\usepackage{array}
\usepackage{amsthm}
\usepackage{multirow}
\usepackage{soul}
\usepackage{nomencl}
\makenomenclature
\title{\LARGE \bf
State of Health Estimation for Battery Modules with Parallel-Connected Cells Under Cell-to-Cell Variations}

\author{Qinan Zhou$^{1,*}$, Dyche Anderson$^{2}$, and Jing Sun$^{3}$% <-this % stops a space
\thanks{$^{1}$Qinan Zhou is with the Department of Mechanical Engineering, University of Michigan, Ann Arbor, MI 48103, USA. Email: {\tt\small qinan@umich.edu}}%
\thanks{$^{2}$Dyche Anderson is with Research and Advanced Engineering, Ford Motor Company, Dearborn, MI 48121, USA. Email: {\tt\small rander34@ford.com}}%
\thanks{$^{3}$Jing Sun is with the Department of Naval Architecture and Marine Engineering, University of Michigan, Ann Arbor, MI 48103, USA. Email: {\tt\small jingsun@umich.edu}}%
\thanks{$^{*}$Corresponding Author.}%
}

\begin{document}

\maketitle
\thispagestyle{empty}
\pagestyle{empty}
%%%%%%%%%%%%%%%%%%%%%%%%%%%%%%%%%%%%%%%%%%%%%%%%%%%%%%%%%%%%%%%%%%%%%%%%%%%%%%%%
\begin{abstract}
State of health (SOH) estimation for lithium-ion battery modules with cells connected in parallel is a challenging problem, especially with cell-to-cell variations. Incremental capacity analysis (ICA) and differential voltage analysis (DVA) are effective at the cell level, but a generalizable method to extend them to module-level SOH estimation remains missing, when only module-level measurements are available. This paper proposes a new method and demonstrates that, with multiple features systematically selected from the module-level ICA and DVA, the module-level SOH can be estimated with high accuracy and confidence in the presence of cell-to-cell variations. First, an information theory-based feature selection algorithm is proposed to find an optimal set of features for module-level SOH estimation. Second, a relevance vector regression (RVR)-based module-level SOH estimation model is proposed to provide both point estimates and three-sigma credible intervals while maintaining model sparsity. With more selected features incorporated, the proposed method achieves better estimation accuracy and higher confidence at the expense of higher model complexity. When applied to a large experimental dataset, the proposed method and the resulting sparse model lead to module-level SOH estimates with a 0.5\% root-mean-square error and a 1.5\% average three-sigma value. With all the training processes completed offboard, the proposed method has low computational complexity for onboard implementations.
\end{abstract}

%%%%%%%%%%%%%%%%%%%%%%%%%%%%%%%%%%%%%%%%%%%%%%%%%%%%%%%%%%%%%%%%%%%%%%%%%%%%%%%%
\nomenclature[01]{$\mathcal{A}$}{Set of All Features}
\nomenclature[02]{$F, G, H$}{Random Variable}
\nomenclature[03]{$I$}{Mutual Information or Conditional Mutual Information}
\nomenclature[04]{$\tilde{I}$}{Normalized Mutual Information or Normalized Conditional Mutual Information}
\nomenclature[05]{$i, j, l$}{Index}
\nomenclature[06]{$K$}{Kernel Function}
\nomenclature[07]{$N$}{Total Number of Sample Points}
\nomenclature[08]{$Q_c$}{Charged Capacity}
\nomenclature[09]{$\mathcal{R}$}{Removed Feature Set}
\nomenclature[10]{$\mathcal{S}$}{Ranked Selected Feature Set}
\nomenclature[11]{$\mathcal{U}$}{Set of Features Not Selected Yet}
\nomenclature[12]{$V$}{Voltage}
\nomenclature[13]{$\boldsymbol{w}$}{Weight Vector}
\nomenclature[14]{$X$}{Feature Random Variable}
\nomenclature[15]{$\boldsymbol{x}$}{Input Vector}
\nomenclature[16]{$Y$}{Output Random Variable}
\nomenclature[17]{$\boldsymbol{y}$}{Vector Containing All Outputs in the Dataset}
\nomenclature[18]{$Z$}{Any Random Variables or Combinations of Random Variables}
\nomenclature[19]{$\boldsymbol{\alpha}$}{Vector with $\alpha_i$ where $\alpha_i^{-1}=$ Variance of $i$-th Weight Distribution}
\nomenclature[20]{$\beta$}{Scalar where $\beta^{-1} =$ Variance of Noise Distribution}
\nomenclature[21]{$\boldsymbol{\mu}$}{Mean Vector}
\nomenclature[22]{$\boldsymbol{\Phi}$}{Kernel Matrix}
\nomenclature[23]{$\boldsymbol{\Sigma}$}{Covariance Matrix}
\printnomenclature
%%%%%%%%%%%%%%%%%%%%%%%%%%%%%%%%%%%%%%%%%%%%%%%%%%%%%%%%%%%%%%%%%%%%%%%%%%%%%%%%
\section{INTRODUCTION}
\label{Intro}
Lithium-ion battery state of health (SOH) estimation with onboard measurements is an active research topic for electrified vehicles \cite{Barre}. SOH, representing the degradation status of batteries, is critical information for battery management systems \cite{Noura}. Depending on the applications, SOH may be defined as resistance rising, capacity fading, or their combinations \cite{Berecibar,CalRegulation}. This paper focuses on capacity fading, and degradation in SOH results in the range reduction of electrified vehicles \cite{CalRegulation}.

While there is a rich set of methodologies and results in the literature for SOH estimation, most of the work reported focuses on the cell level \cite{Noura,Berecibar,Yao}, using model-based and data-driven methods. Popular model-based methods include Bayesian filters \cite{PlettEKF,PlettSPKF,Schwunk}, sliding-mode observers \cite{Kim}, and least-square techniques \cite{PlettLS}. However, given complicated degradation processes, robust and accurate estimation is difficult to achieve with simple phenomenological or physical models using only onboard measurements. On the other hand, data-driven methods have the advantage of not requiring physics-based models. Various data-driven methods include convolutional and recurrent neural networks \cite{Yang,Chaoui,Gao2}. Without insights and structures from physics-based models, however, the effectiveness of these methods solely relies on the comprehensiveness and quality of training data.

\begin{figure}[b]
    \centering
    \begin{subfigure}[h]{0.23\textwidth}
    \centering
    \includegraphics[width=\textwidth,trim=0.1 0.1 25 25,clip]{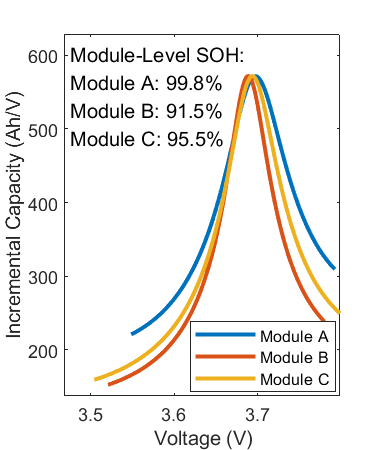}
    \caption{Diff. SOH, Same IC Peak}
    \label{fig:Diff_SOH_Same_IC}
    \end{subfigure} %\hspace{0.7cm}
    \begin{subfigure}[h]{0.23\textwidth}
    \centering
    \includegraphics[width=\textwidth,trim=0.1 0.1 25 25,clip]{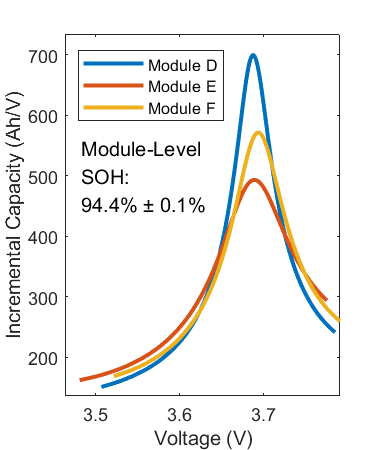}
    \caption{Same SOH, Diff. IC Peak}
    \label{fig:Same_SOH_Diff_IC}
    \end{subfigure}
\caption{Example Module-Level IC Curves around an IC Peak in the Presence of Cell-to-Cell Variations}
\label{fig:IC_Curve_C2Cvar}
\end{figure} 

One promising middle ground between model-based and data-driven methods is incremental capacity analysis (ICA) \cite{Weng1,Weng2,Weng3}. Peaks in incremental capacity (IC) curves reflect degradation mechanisms inside batteries \cite{Dubarry,Krupp}. Thus, ICA incorporates aging physics without using complicated mechanistic models. The capability to provide physical insights about degradation is the key strength of ICA over deep neural network-based methods and other data-based approaches. Furthermore, this method leverages data-driven techniques to correlate physical features (IC peaks) and SOH, and uses the learned model to estimate SOH \cite{Zhou1}. ICA has been shown to be effective for different battery chemistries under different temperatures, C rates, and charging ranges \cite{Zhou1}. Differential voltage analysis (DVA), a similar method, has also been demonstrated to be effective for cell-level SOH estimation \cite{Wang}.

\begin{figure*}[ht]
    \centering
    \includegraphics[width=0.99\textwidth,trim=2 5 2 2,clip]{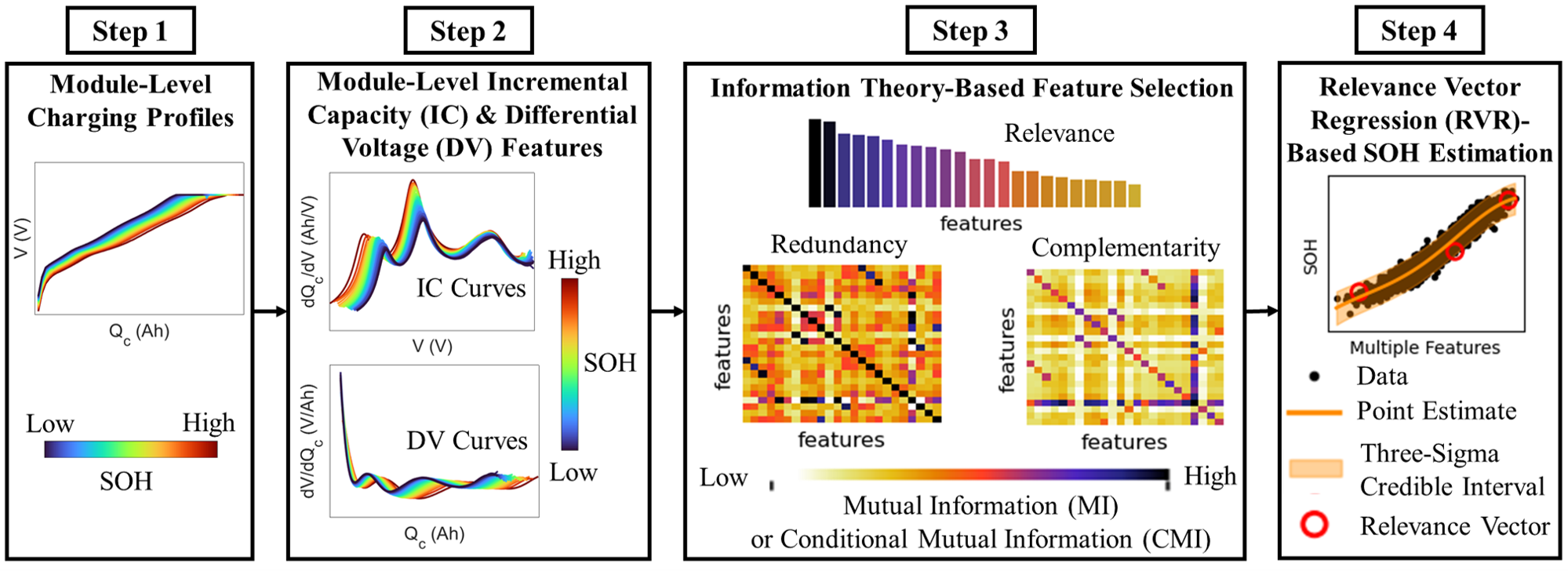}
    \caption{Overview of Proposed Module-Level State of Health Estimation Method}
    \label{fig:Algorithm}
\end{figure*} 

Battery packs consist of multiple cells in parallel or series for high-energy and high-power applications \cite{PlettBook}. In this paper, a battery module refers to a set of battery cells connected in parallel, and the module-level SOH is defined as:
\begin{equation}
    \text{SOH} = \frac{Q}{Q_{\text{fresh}}},
\end{equation}
where $Q$ and $Q_{\text{fresh}}$ are the capacities of a module at its current and fresh state, respectively. While this definition of the module-level SOH has the advantage of directly relating SOH to the module-level capacity, it should be noted that it does not capture cell-to-cell variations within modules. For example, a 95\% SOH module can consist of three cells all having 95\% cell-level SOHs or three cells with 100\%, 95\%, and 90\% cell-level SOHs, respectively.

For battery modules with cells connected in parallel, inevitable cell-to-cell variations within modules exist in cell capacities, internal resistances, contact resistances, temperatures, etc. \cite{Gong,Baumann}. Different papers in the literature have reported contradictory results about the evolution of cell-to-cell variations within modules over time. Take cell-to-cell variations in capacities and internal resistances as examples. \cite{Brand} and \cite{Fernández} have reported that capacities and internal resistances of individual cells within modules converge due to self-balancing, whereas \cite{Gong} and \cite{Baumann} have reported diverging results. Thus, unless the research community comes to a solid conclusion about how cell-to-cell variations within modules will evolve, the module-level SOH estimation must be performed under the assumption that cell-to-cell variations exist within battery modules.

For module-level SOH estimation, however, cell-to-cell variations within modules hinder ICA- and DVA-based methods \cite{An}. Cell-to-cell variations cause uneven current distribution among individual cells and, thus, distort the module-level IC and DV features \cite{Weng4}. Take IC peak, a feature used for cell-level SOH estimation \cite{Weng1,Weng2,Weng3,Zhou1}, as an example. For lithium nickel-manganese-cobalt oxide 622 (NMC622) modules with three cells connected in parallel (208Ah nominal module capacity), experimental data show that the module-level IC peak could be the same for modules with different module-level SOH (Fig. \ref{fig:Diff_SOH_Same_IC}) or different for modules with the same module-level SOH (Fig. \ref{fig:Same_SOH_Diff_IC}) under 86A constant-current charging and room temperature. Thus, the correlation between IC peaks and SOH is obscured by cell-to-cell variations. Some features could be found robust to cell-to-cell variations, but these are only phenomenological observations and the insensitivity of these features could disappear when battery chemistries or charging conditions change. For example, IC peaks and distances between two DV inflection points are found robust to cell-to-cell variations in \cite{Weng4} and \cite{Wang2}, respectively, for lithium iron phosphate (LFP) modules under their specific conditions. However, for the NMC622 modules discussed above, the IC peak changes significantly due to cell-to-cell variations, while only one DV inflection point exists at the module level. Thus, a generalizable method to find optimal module-level IC and DV features for module-level SOH estimation is still missing.

In this paper, a new and generalizable method, as shown in Fig. \ref{fig:Algorithm}, is proposed to estimate module-level SOH with high accuracy and confidence in the presence of cell-to-cell variations. Instead of selecting features based on phenomenological observations as in the existing approaches, this paper proposes an algorithm that can find optimal features for module-level SOH estimation. The optimal features can change when situations change, but the proposed algorithm stays the same. Thus, the proposed algorithm is generalizable. Specifically, the contributions of the paper are three-fold: 
\begin{itemize}
    \item First, an information theory-based feature selection algorithm is proposed to find an optimal set of features for module-level SOH estimation. The feature selection process is independent of subsequent learning algorithms. To the best of our knowledge, it is the first application of the information theory-based feature selection to module-level SOH estimation. 
    \item Second, a relevance vector regression (RVR)-based SOH estimation model is developed to not only extract point estimates for SOH but also obtain three-sigma credible intervals. The sparsity of the model is automatically imposed, which leads to low model complexity and makes it suitable for onboard implementations.
    \item Third, applied to a large dataset of NMC622 modules, the proposed method achieves a good module-level SOH estimation performance (0.5\% root-mean-square error (RMSE) and 1.5\% average three-sigma value) with a low computational requirement, making it feasible for onboard implementations.
\end{itemize}

To elucidate the proposed method in Fig. \ref{fig:Algorithm}, the paper is organized as follows. Section \ref{Features} summarizes all the IC and DV features related to SOH in general. Section \ref{InfoFS} develops the proposed information theory-based feature selection algorithm. Section \ref{RVR} presents the proposed RVR-based SOH estimation model. Section \ref{Dataset} discusses the battery datasets used in this study. Section \ref{Cell_Result} demonstrates how to interpret the proposed method using experimental battery cell data. Section \ref{Results} assesses the performance of the proposed method using experimental battery module data and discusses the onboard computational requirements involved with the implementation of the proposed method. Section \ref{Conclusion} summarizes the paper.

%%%%%%%%%%%%%%%%%%%%%%%%%%%%%%%%%%%%%%%%%%%%%%%%%%%%o%%%%%%%%%%%%%%%%%%%%%%%%%%%%
\section{BATTERY MODULE-LEVEL IC AND DV FEATURES}
\label{Features}

\begin{figure*}[ht]
    \centering
    \begin{subfigure}[h]{0.3\textwidth}
        \centering
        \includegraphics[width=\textwidth,trim=10 1 40 25,clip]{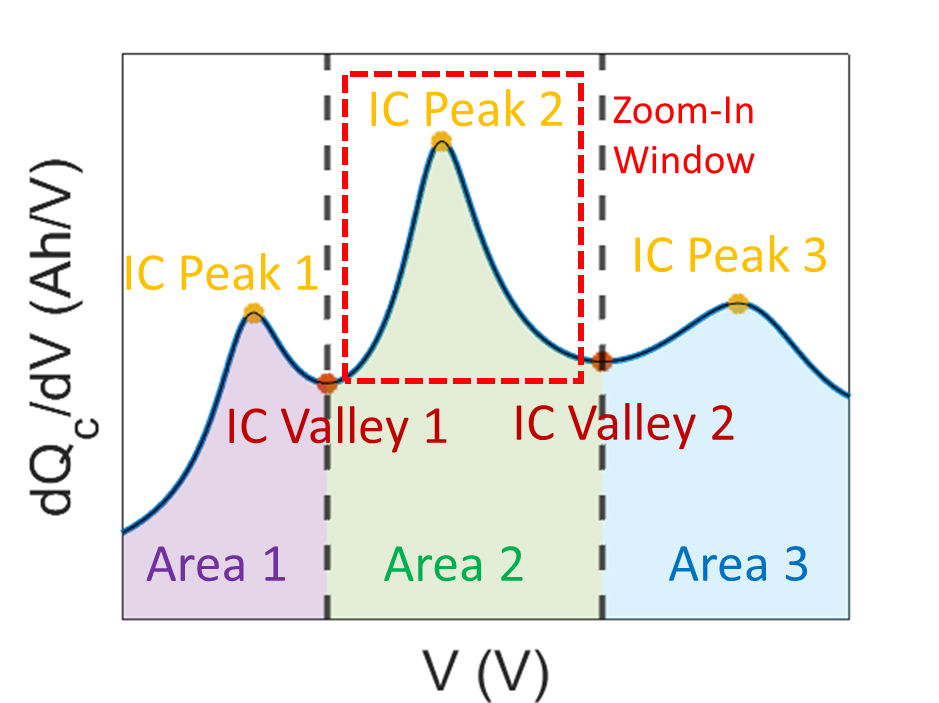}
        \caption{IC Curve}
        \label{fig:IC_Curve_F1}
    \end{subfigure} \hfill
    \begin{subfigure}[h]{0.3\textwidth}
        \centering
        \includegraphics[width=\textwidth,trim=10 1 40 25,clip]{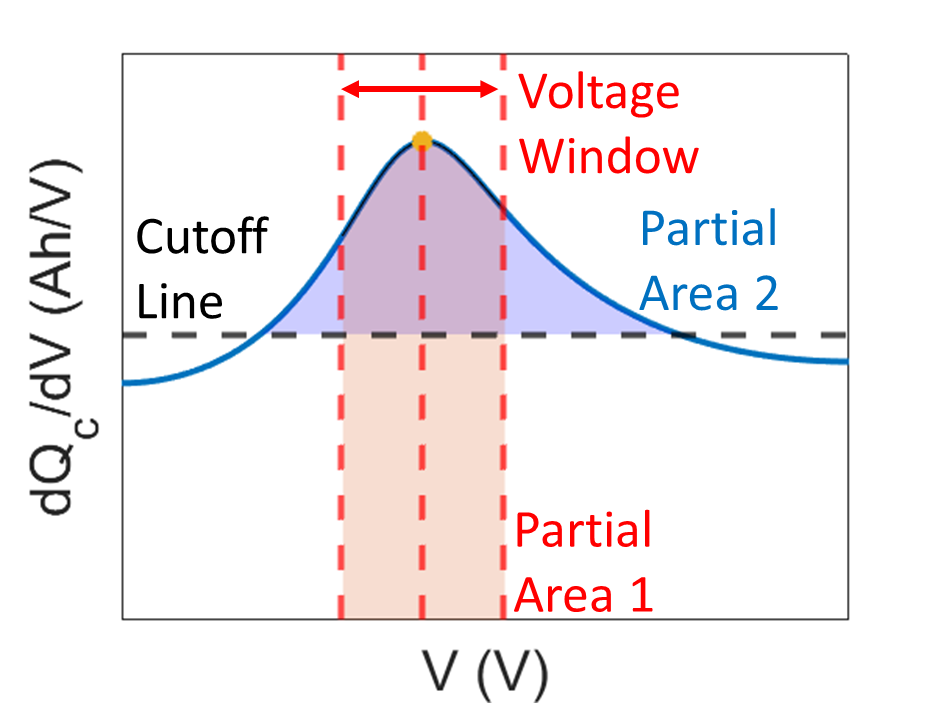}
        \caption{Zoomed-In IC Curve at Example Peak}
        \label{fig:Zoomed_In_IC_Curve_F2}
    \end{subfigure} \hfill
    \begin{subfigure}[h]{0.3\textwidth}
        \centering
        \includegraphics[width=\textwidth,trim=10 1 40 25,clip]{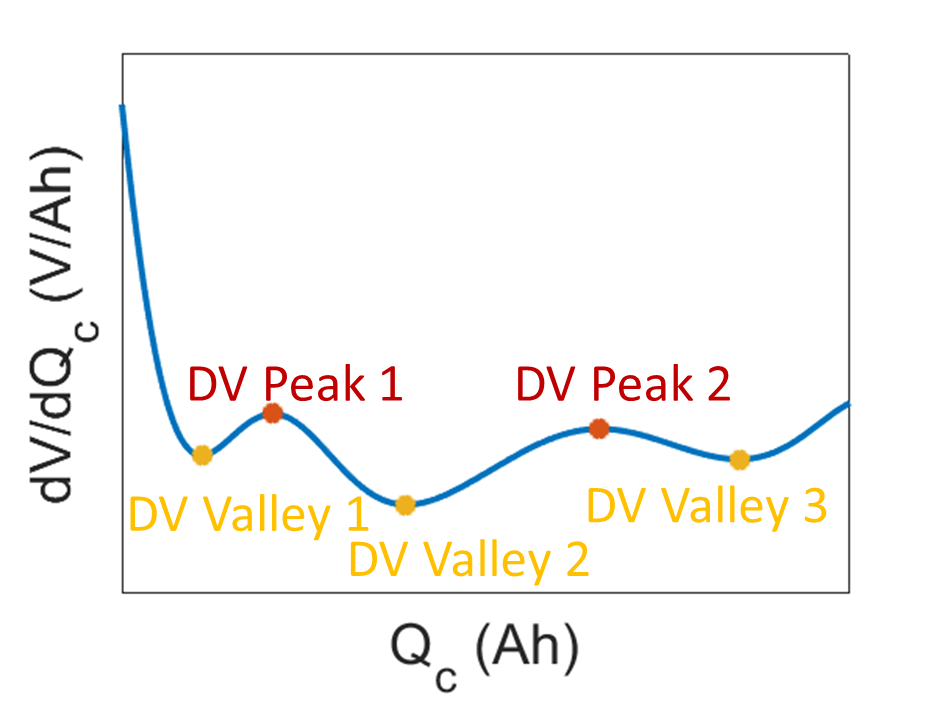}
        \caption{DV Curve}
        \label{fig:DV_Curve_F3}
    \end{subfigure} 
\caption{Definitions of Incremental Capacity (IC) and Differential Voltage (DV) Features}
\label{fig:Feature_Def}
\end{figure*}

This section deals with Step 2 of Fig. \ref{fig:Algorithm} where all the available features in module-level IC and DV curves are extracted. Given measured module-level charged capacity-voltage ($Q_c$-$V$) profiles under constant-current regimes, the module-level IC and DV are defined as $\text{IC} = dQ_c/dV$ and $\text{DV} = dV/dQ_c$, respectively. In this study, support vector regression is used to fit the measured profiles and, then, module-level IC and DV curves are obtained through analytical derivative calculations to mitigate noise sensitivity caused by numerical derivatives \cite{Weng1,Weng2,Weng3}.

The following features, as defined graphically in Fig. \ref{fig:Feature_Def}, can be extracted from the IC and DV curves. For IC features, as shown in Fig. \ref{fig:IC_Curve_F1}, the $x$- and $y$-coordinates of IC peaks and valleys are their locations and heights, respectively. IC peak areas 1, 2, and 3 are the areas from the minimum voltage to the first IC valley location, from the first to the second IC valley locations, and from the second IC valley location to the maximum voltage, respectively \cite{Anseán}. IC peak partial areas can be defined as the area either above a user-defined horizontal cutoff line or within a user-defined symmetric voltage window around a target IC peak \cite{ZhouR}, as shown in Fig. \ref{fig:Zoomed_In_IC_Curve_F2}. Similarly, DV features, as defined in Fig. \ref{fig:DV_Curve_F3}, include the locations ($x$-coordinates) and heights ($y$-coordinates) of DV peaks and valleys. For charging condition features, temperatures and C rates of constant-current charging regimes affect IC and DV curves. Thus, they are also important features for SOH estimation \cite{Zhou1}.

Table \ref{table:Features} summarizes all the available features from ICA, DVA, and charging conditions. Their physical meanings and relations to battery degradation are discussed in the references in Table \ref{table:Features}. Note that, depending on battery chemistries and charging conditions, some features in Table \ref{table:Features} might not be available in practice, resulting in different numbers of features for different datasets (as to be seen in Section \ref{Dataset}).

\begin{table}[H]
\centering
\caption{Summary of Features from ICA, DVA, and Charging Conditions}
\label{table:Features}
\begin{tabular}{|c|c|c|c|}
\hline
\textbf{Feature Category} & \textbf{Acronym} & \textbf{Reference} \\ \hline\hline

IC Peak Height {[}Ah/V{]} & IC PH & \multirow{8}{*}{\begin{tabular}[c]{@{}c@{}} \cite{Dubarry} \\ \cite{Krupp} \\ \cite{Bloom} \end{tabular} }\\ \cline{1-2}

IC Peak Location {[}V{]} & IC PL &  \\ \cline{1-2}

DV Valley Height {[}V/Ah{]} & DV VH & \\ \cline{1-2}

DV Valley Location {[}Ah{]} & DV VL & \\ \cline{1-2}

IC Valley Height {[}Ah/V{]} & IC VH &  \\ \cline{1-2}

IC Valley Location {[}V{]} & IC VL & \\ \cline{1-2}

DV Peak Height {[}V/Ah{]} & DV PH & \\ \cline{1-2}

DV Peak Location {[}Ah{]} & DV PL & \\ \cline{1-3}  

IC Peak Area {[}Ah{]} & IC AR & \begin{tabular}[c]{@{}c@{}} \cite{Anseán} \end{tabular}\\ \cline{1-3} \cline{3-3} 

IC Peak Partial Area {[}Ah{]} & IC PA & \begin{tabular}[c]{@{}c@{}} \cite{ZhouR} \end{tabular} \\ \hline

Temperature {[}°C{]} & - & \multirow{2}{*}{\begin{tabular}[c]{@{}c@{}} \cite{Zhou1} \end{tabular}} \\ \cline{1-2}

C Rate {[}C{]} & - &  \\ \hline
\end{tabular}
\end{table}
%%%%%%%%%%%%%%%%%%%%%%%%%%%%%%%%%%%%%%%%%%%%%%%%%%%%%%%%%%%%%%%%%%%%%%%%%%%%%%%%
\section{INFORMATION THEORY-BASED FEATURE SELECTION}
\label{InfoFS}
The total number of features that can be extracted from ICA, DVA, and charging conditions could be very large. It is necessary to select a subset of features for SOH estimation and avoid the ``curse of dimensionality" \cite{Li2}. This section corresponds to Step 3 of Fig. \ref{fig:Algorithm} where an algorithm is proposed to find optimal sets of features for SOH estimation.

Compared to other approaches, the information theory-based feature selection approach is adopted in this work for its excellent capability to capture nonlinear relationships and interactions among multiple features and targets, abundance of feature selection criteria established in the literature, and computational efficiency \cite{Li2}. In the context of module-level SOH estimation, this approach leverages data to infer underlying physical insights in the presence of cell-to-cell variations. Moreover, its results are independent of subsequent SOH estimation algorithms \cite{Li2}, providing the flexibility for adopting different methods in Step 4 of Fig. \ref{fig:Algorithm} for the final SOH estimation. 

\subsection{Definitions}
\label{InfoDefn}
To present the proposed algorithm, the related concepts and terminologies are defined first. Mutual information (MI) and conditional mutual information (CMI) are two key concepts for information theory-based feature selection \cite{Li2}. Most generally, they are defined among a mixture of continuous and discrete random variables (RV) using measure theory \cite{Gao,Mesner}. Considering three discrete RVs $F$, $G$, and $H$, the MI between $F$ and $G$ is defined as:
\begin{equation}
    I\left(F; G\right) = \sum_{f \in F} \sum_{g \in G} p\left(f,g\right) \log{ \frac{p\left(f,g\right)}{p\left(f\right)p\left(g\right)} }, \label{eqn:MI_def}
\end{equation}
and the CMI between $F$ and $G$ given $H$ are defined as: 
\begin{align}
I\left(F; G | H\right)  
&= \sum_{h \in H} \biggl\{ p\left(h\right) \cdot \notag \\& \sum_{f \in F} \sum_{g \in G} p\left(f,g|h\right) \log{ \frac{p\left(f,g|h\right)}{p\left(f|h\right)p\left(g|h\right)} }  \biggl\}, \label{eqn:CMI_def}
\end{align}
where $p(\cdot)$ is the probability distribution function \cite{InfoBook}. If $F$, $G$, and $H$ are continuous RVs, summations in these definitions are replaced by integrations \cite{Kraskov}. Here, natural logarithms are used. Intuitively, the MI measures the uncertainty reduction of an RV if another RV is known, while the CMI measures the uncertainty reduction of an RV if another RV is known given the third RV \cite{InfoBook}.

When uncertainty itself varies greatly from RV to RV, MI (uncertainty reduction) and CMI (conditional uncertainty reduction) become ambiguous \cite{Estevez}. Thus, they should be normalized \cite{Estevez}. This paper uses the following normalized MI and CMI \cite{Kvålseth}:   
\begin{eqnarray}
\tilde{I}\left(F;G\right) &=& \frac{I\left(F;G\right)}{\min \left(I\left(F;F\right),I\left(G;G\right)\right)}, \label{eqn:MI_norm}\\ 
\tilde{I}\left(F;G | H\right) &=& \frac{I\left(F;G | H \right)}{\min \left(I\left(F;F\right),I\left(G;G\right)\right)}. \label{eqn:CMI_norm}
\end{eqnarray}
Normalization (\ref{eqn:MI_norm}) is chosen to keep the symmetry property of the unnormalized MI \cite{Kvålseth} and provide a number in $\left[0,1\right]$, similar to a correlation coefficient \cite{Estevez}. Note that Normalization (\ref{eqn:CMI_norm}) is with respect to $\min \left(I\left(F;F\right),I\left(G;G\right)\right)$, instead of $\min \left(I\left(F;F|H\right),I\left(G;G|H\right)\right)$. This is a decision made for the feature selection of SOH estimation problems to avoid numerical issues, as to be discussed in Remark \ref{remark_on_CMI_norm}, Section \ref{Cell_Result}. 

\subsection{Feature Selection Algorithm}
\label{FS_Algm}

\begin{algorithm*}[h]
\textbf{Output}: ranked selected feature set $\mathcal{S}$, unranked removed feature set $\mathcal{R}$ \

\textbf{Input}: set $\mathcal{A}$ containing all the features, threshold $\tilde{I}_{\text{th}}$ for removing completely redundant features\

 \textbf{Initialization}: $\mathcal{S} \leftarrow \emptyset$ or \{pre-selected features\}, $\mathcal{R} \leftarrow \emptyset$, $\mathcal{U}  \leftarrow \mathcal{A} \backslash \mathcal{S} $\

 \

 Find completely redundant features to any pre-selected features: $\mathcal{D} \leftarrow \left\{ X \in \mathcal{U}: \tilde{I}\left( X; X_j \right) \geq \tilde{I}_{\text{th}}, X_j \in \mathcal{S} \right\}$\
 
 $\mathcal{U} \leftarrow \mathcal{U} \backslash \mathcal{D}$, $\mathcal{R}\leftarrow \mathcal{R} \cup \mathcal{D}$ \
 
 \While{$|\mathcal{U}| > 0$}{

 \If{$ |\mathcal{S}| = 0$}{
 $ X^* \leftarrow \operatorname*{argmax}_{X \in \mathcal{U} } \tilde{I}\left(X ; Y \right)$   \
 }
 \Else{
 $ X^* \leftarrow \operatorname*{argmax}_{X \in \mathcal{U} } \left( \tilde{I}\left( X; Y \right) - \frac{1}{|\mathcal{S}|} \sum_{X_j \in \mathcal{S}} \tilde{I} \left(X ; X_j \right) + \frac{1}{|\mathcal{S}|} \sum_{X_j \in \mathcal{S}} \tilde{I} \left(X; X_j | Y \right) \right)$ \
 }
 $\mathcal{S} \leftarrow \mathcal{S} \cup \{ X^* \}$, $\mathcal{U} \leftarrow \mathcal{U} \backslash \{ X^* \}$ \

 Find completely redundant features to $X^*$: $\mathcal{D} \leftarrow \left\{ X \in \mathcal{U} : \tilde{I} \left(X^*; X \right) \geq \tilde{I}_{\text{th}} \right\}$\

 $\mathcal{U}\leftarrow \mathcal{U} \backslash \mathcal{D}$, $\mathcal{R}\leftarrow \mathcal{R} \cup \mathcal{D}$ \
 
 }
\caption{Proposed Information Theory-Based Feature Selection Algorithm}
\label{alg:infoFS}
\end{algorithm*}

In general, searching for the globally optimal set of features is NP-hard and heuristic sequential search routines could be used \cite{Li2}. The widely-used forward search algorithm is adopted in this paper, because of its computational efficiency \cite{Ferri}. It adds new features to a set $\mathcal{S}$ one by one according to a feature selection criterion, which will be discussed later in this subsection. The order in which features are added reflects their relative importance to SOH estimation. Thus, the output $\mathcal{S}$ is a ranked set. The ranking can be used to select a proper subset of $\mathcal{S}$ for SOH monitoring with a relatively simple model. Note that the selected features could be sub-optimal due to the greedy nature of the search algorithm \cite{Li2}.

The pseudo-code for the proposed feature selection algorithm is given in Algorithm \ref{alg:infoFS}. Symbols used in Algorithm \ref{alg:infoFS} are defined as: $\mathcal{A}$ -- the set of all features, $\mathcal{S}$ -- ranked selected feature set, $\mathcal{U}$ -- the set of features not selected yet, $\mathcal{R}$ -- removed feature set. $\mathcal{S}$ and $\mathcal{R}$ are the outputs of Algorithm \ref{alg:infoFS}.

Starting with $\mathcal{S} = \emptyset$, features are added into $\mathcal{S}$ iteratively according to Algorithm \ref{alg:infoFS}. At the $l$-th iteration, given $\mathcal{S}_{l-1}$ and $\mathcal{U}_{l-1}$ resulted from the previous iteration, the algorithm solves the following optimization problem:
\begin{equation}
    X^*_l = \operatorname*{argmax}_{X \in \mathcal{U}_{l-1} } J\left(X \right), \label{eqn:FS_Optimization}
\end{equation}
where the feature selection criterion $J\left(X \right)$ is modified from the joint mutual information criterion \cite{Meyer} and defined as: 
\begin{multline}
    J\left(X \right) = \tilde{I}\left( X; Y \right) - \frac{1}{|\mathcal{S}_{l-1}|} \sum_{X_j \in \mathcal{S}_{l-1}} \tilde{I} \left(X ; X_j \right) \\ + \frac{1}{|\mathcal{S}_{l-1}|} \sum_{X_j \in \mathcal{S}_{l-1}} \tilde{I} \left(X; X_j | Y \right),
    \label{eqn:FS_obj}
\end{multline}
where $|\cdot|$ is the cardinality of a set, $X \in \mathcal{U}_{l-1}$ is a feature not selected yet, $X_j \in \mathcal{S}_{l-1}$ are selected features ($j$ is an index), and $Y$ is the output (i.e., SOH). $J(X)$ contains three terms, namely the relevance $\tilde{I}\left( X; Y \right)$, the average redundancy $1/|\mathcal{S}_{l-1}| \cdot \sum_{X_j \in \mathcal{S}_{l-1}} \tilde{I} \left(X ; X_j \right)$, and the average complementarity $1/|\mathcal{S}_{l-1}| \cdot \sum_{X_j \in \mathcal{S}_{l-1}} \tilde{I} \left(X; X_j | Y \right)$. These terms are explained as follows: 
\begin{itemize}
    \item \textbf{Relevance}: The relevance of $X$ to $Y$ is defined as the normalized MI between them, i.e., $\tilde{I}\left( X; Y \right)$. The higher the relevance is, the more information is shared between $X$ and $Y$. For cell-level SOH estimation, maximizing the relevance picks up features physically related to SOH. For module-level SOH estimation, features with high relevance are physically related to SOH and have small sensitivity to cell-to-cell variations.  
    \item \textbf{Redundancy}: The redundancy between $X$ and $X_j$ is defined as the normalized MI between them, i.e., $\tilde{I} \left(X; X_j \right)$. A high redundancy implies that $X$ and $X_j$ have a lot in common, so there is no need to keep both. $ \sum_{X_j \in \mathcal{S}_{l-1}} \tilde{I} \left(X ; X_j \right)$ gives the total redundancy between $X$ and all the selected features in $\mathcal{S}_{l-1}$. Since this total redundancy grows very fast as the cardinality of $\mathcal{S}_{l-1}$ increases through iterations, the factor $1/|\mathcal{S}_{l-1}|$ is multiplied to bring down the scale \cite{Meyer}. The proposed algorithm minimizes redundancy. 
    \item \textbf{Complementarity}: The complementarity captures the ``synergy" between two features \cite{Vergara}. Some features could have low relevance to $Y$ but become important in the context of other features. For example, the C rate does not show strong relevance to SOH, because SOH cannot be determined given the C rate alone. However, for example, when IC peaks are chosen, the C rate becomes highly important, as the C rate affects IC peaks \cite{Zhou1}. The complementarity between $X$ and $X_j$ is defined as the normalized CMI between them given $Y$, i.e., $\tilde{I} \left(X ; X_j | Y\right)$. Similarly, $ \sum_{X_j \in \mathcal{S}_{l-1}} \tilde{I} \left(X ; X_j |Y\right)$ is the total complementarity, while $1/|\mathcal{S}_{l-1}|$ provides the scaling effect \cite{Meyer}. By maximizing the complementarity, the proposed algorithm leverages the ``synergy" between the newly selected feature and those in $\mathcal{S}_{l-1}$.
\end{itemize}
Thus, for module-level SOH estimation, at each iteration, the proposed algorithm selects the feature based on whether it provides the best trade-off among relevance, redundancy, and complementarity, not on how much it is distorted by cell-to-cell variations. Optimization (\ref{eqn:FS_Optimization}) is performed until $|\mathcal{U}| =0$.
\begin{remark}
    (On Non-Empty Initialization) $\mathcal{S}$ can also be initialized to contain some pre-selected features based on available physical knowledge. Note that these pre-selected features can be unranked or ranked. For this study, $\mathcal{S}$ is initialized as an empty set to let data speak for themselves. 
\end{remark}
\begin{remark}
    (On Removing Features) For SOH estimation problems, normalized CMI could be high for both complementary and completely redundant features, causing the latter to be selected. Thus, the feature removal step in Algorithm \ref{alg:infoFS} is necessary. Specifically, at one iteration, if $X^*$ is selected by solving Optimization (\ref{eqn:FS_Optimization}), then the proposed algorithm removes all the features $X \in \mathcal{U}$ whose redundancies $\tilde{I} \left(X^* ; X \right)$ are above a user-defined threshold $\tilde{I}_{\text{th}}$, i.e., removes the set $\mathcal{D} = \left\{ X \in \mathcal{U} : \tilde{I} \left(X^*; X \right) \geq \tilde{I}_{\text{th}} \right\}$ from $\mathcal{U}$. Thus, these features will not be considered in the next iteration.
\end{remark}

\subsection{MI/CMI Estimation and Data Standardization}
\label{CMI_Est}
The proposed algorithm relies on MI and CMI whose definitions involve probability distributions of RVs. However, these distributions are not known in practice. Thus, MI and CMI must be estimated using data.

To have consistent estimation results, this paper uses the $k$-nearest neighbor ($k$NN) CMI estimator \cite{Mesner} to calculate both MI and CMI. Consider sample $\left\{\left( f_i,g_i,h_i \right)^T\right\}_{i=1}^N$ from RVs $\left( F, G, H \right)^T$ where $N$ is the number of sample points and $i$ is an index. For any sample point $\left( f_i,g_i,h_i \right)^T$, one can first define its $k$-nearest neighbor based on $l_\infty$-norm distance and denote the corresponding $l_\infty$-norm distance as $\frac{1}{2} d_{k,i}$. Note that $k$ is a tuning hyperparameter for the estimator and its proper value is related to the number of data \cite{Kraskov}. Next, one can count the number of sample points within different windows: 
\begin{equation}
    n_{Z,i} = \left| \left\{ z_j : \left\| z_i - z_j \right\|_\infty \leq \frac{1}{2} d_{k,i}, i \neq j\right\} \right|, \label{eqn:CountKNN}
\end{equation}
where $Z$ can be any RVs or combinations of RVs (e.g., $\left(F,H\right)^T$, $\left(G,H\right)^T$, $H$, etc.), and $z$ can be any sample points from these RVs or combinations of RVs (e.g., $\left(f,h\right)^T$, $\left(g,h\right)^T$, $h$, etc.). Finally, by repeating the same process for all the data, the estimator \cite{Mesner} computes the CMI estimates $\hat{I}\left(F;G|H\right)$ as:  
\begin{eqnarray}
    \xi_i = \psi\left( \tilde{k}_i \right) - \psi\left( n_{FH,i}\right) - \psi\left( n_{GH,i}\right) + \psi\left( n_{H,i}\right), \label{eqn:CMI_estimate_1} \\
    \hat{I}\left(F;G|H\right) = \max \left\{ \frac{1}{N} \sum_{i=1}^N \xi_i,0\right\}, \label{eqn:CMI_estimate_2}
\end{eqnarray}
where $\tilde{k}_i$ is calculated using Equation (\ref{eqn:CountKNN}) with $Z$ and $z$ defined as $\left(F,G,H\right)^T$ and $\left(f,g,h\right)^T$, $\psi \left( \cdot \right)$ is the digamma function that satisfies $\psi \left( x+1 \right) = \psi \left( x \right)+1/x$, $\psi \left( 1 \right) = -C$, and $C$ is the Euler-Mascheroni constant \cite{Kraskov}. 

To estimate MI $\hat{I}\left(F;G\right)$, one can still use Equations (\ref{eqn:CMI_estimate_1}) and (\ref{eqn:CMI_estimate_2}) by setting $H$ to be an artificially generated white Gaussian noise with a mean of 0 and variance of 1. The white Gaussian noise is generated such that it is independent of $F$, $G$, and $\left(F,G\right)^T$ jointly. Thus, based on Definitions (\ref{eqn:MI_def}) and (\ref{eqn:CMI_def}), 
\begin{equation}
    \hat{I}\left(F;G\right) = \hat{I}\left(F;G| \text{White Gaussian Noise} \right).
\end{equation}

When $F$, $G$, and $H$ have different orders of magnitudes (as in the case of IC and DV features), the use of $l_\infty$-norm in MI and CMI estimation causes numerical problems. A common solution given in the literature is to standardize all the sample points first using:
\begin{equation}
    z_i^{(s)} = \frac{z_i - \text{mean}\left( \left\{ z_i \right\}_{i=1}^N \right)}{\text{std}\left( \left\{ z_i \right\}_{i=1}^N \right)} \label{eqn:standardize}
\end{equation}
for sample $\left\{ z_i \right\}_{i=1}^N$ from any RV $Z$, where $\text{mean}\left( \cdot \right)$ and $\text{std}\left( \cdot \right)$ are the mean and standard deviation of sample points, respectively \cite{Runge}.
%%%%%%%%%%%%%%%%%%%%%%%%%%%%%%%%%%%%%%%%%%%%%%%%%%%%%%%%%%%%%%%%%%%%%%%%%%%%%%%%
\section{RELEVANCE VECTOR REGRESSION}
\label{RVR}

\begin{algorithm*}[h]
\textbf{Output}: relevance vectors $\{\textbf{rv}_j\}_{j=1}^{N_\text{rv}}$, reciprocals of variances for offset and weights $\boldsymbol{\alpha_\text{MP}}$, noise variance $\beta_\text{MP}^{-1}$, trained/pruned $\boldsymbol{\Sigma}$, trained/pruned $\boldsymbol{\mu}$, a boolean variable $o$ about whether offset is used \

\textbf{Input}: standardized data $\left\{ \left( \boldsymbol{x}_i, y_i \right) \right\}_{i=1}^N$, kernel function $K(\cdot, \cdot)$ \

 \textbf{Initialization}: $\boldsymbol{\alpha}$ with $\alpha_i \leftarrow \frac{1}{\left(N+1\right)^2}, i=0,...,N$, $\boldsymbol{\alpha}^{\text{old}}$ with $\alpha^{\text{old}}_i \leftarrow \alpha_i$, $\beta^{-1} \leftarrow \left(0.1 \cdot \text{std}\left(\{y_i\}_{i=1}^N\right)\right)^2$, a set $\mathcal{E} \leftarrow \left\{e_0,\boldsymbol{x}_1,...,\boldsymbol{x}_N \right\}$ ($e_0$ serves as a placeholder for offset and can be initialized as anything), maximum number of iterations $N_{\text{iter}}$, threshold $\alpha_{\text{th}}$ (a large number), tolerance $\text{tol}$, a small number $\varepsilon$, $\boldsymbol{y} = \begin{bmatrix} y_1 & ... & y_N \end{bmatrix}^T$, $\boldsymbol{\Phi} \leftarrow \begin{bmatrix} \boldsymbol{\phi} \left(\boldsymbol{x}_1 \right) & ... & \boldsymbol{\phi} \left(\boldsymbol{x}_N \right) \end{bmatrix}^T \in \mathbb{R}^{N \times \left(N+1\right)}$ where $\boldsymbol{\phi} \left(\boldsymbol{x}_i\right) \leftarrow \begin{bmatrix} 1 & K\left(\boldsymbol{x}_i,\boldsymbol{x}_1\right) & ... & K \left(\boldsymbol{x}_i,\boldsymbol{x}_N\right) \end{bmatrix}^T, \forall i=1,...,N$\

 \
 
 \For{$k = 1,...,N_{\text{iter}}$}{
 
 $\boldsymbol{A}\leftarrow \text{diag}\left( \boldsymbol{\alpha} \right)$, $\boldsymbol{\Sigma} \leftarrow$ Equation (\ref{eqn:SigmaMat}), $\boldsymbol{\mu} \leftarrow$ Equation (\ref{eqn:muVec}) \

 Update $\boldsymbol{\alpha}$ with all the components $\alpha_i \leftarrow$ Equation (\ref{eqn:alpha_update_new}), $\beta^{-1} \leftarrow$ Equation (\ref{eqn:beta_update}) \

 Find indices that need to be kept: $\{i\}^{\text{keep}} \leftarrow \{i | \alpha_i < \alpha_{\text{th}} \}$    

 Prune: $\boldsymbol{\alpha} \leftarrow \boldsymbol{\alpha}\left[ \{i\}^{\text{keep}}\right]$, $\boldsymbol{\alpha}^{\text{old}} \leftarrow \boldsymbol{\alpha}^{\text{old}}\left[ \{i\}^{\text{keep}}\right]$, $\boldsymbol{\Sigma} \leftarrow \boldsymbol{\Sigma}\left[ \{i\}^{\text{keep}},\{i\}^{\text{keep}}\right]$, $\boldsymbol{\mu} \leftarrow \boldsymbol{\mu}\left[ \{i\}^{\text{keep}}\right]$, $\boldsymbol{\Phi} \leftarrow \boldsymbol{\Phi}\left[ :, \{i\}^{\text{keep}}\right]$, $\mathcal{E} \leftarrow \mathcal{E}\left\{ \{i\}^{\text{keep}}\right\} $ \

 \If{$\left\| \boldsymbol{\alpha} - \boldsymbol{\alpha}^{\text{old}}\right\|_\infty \leq \text{tol}$ and $k > 1$}{
 break  \
 }
 
 $ \boldsymbol{\alpha}^{\text{old}} \leftarrow \boldsymbol{\alpha}$ \
 }
 $\boldsymbol{\alpha_{\text{MP}}} \leftarrow \boldsymbol{\alpha}$, $\beta_{\text{MP}}^{-1} \leftarrow \beta^{-1}$ \
 
 $\{\textbf{rv}_j\}_{j=1}^{N_\text{rv}} \leftarrow \begin{cases} \mathcal{E} \text{, if $e_0 \not\in \mathcal{E}$, i.e., the offset is not used} \\ \mathcal{E}\backslash \{e_0\} \text{, otherwise} \end{cases}$, $o \leftarrow \begin{cases} \text{False} \text{, if the offset is not used} \\ \text{True} \text{, otherwise} \end{cases}$ \

 \ 
 
 \tcc{":" means all the rows in matrix $\boldsymbol{\Phi}$} 
 
\caption{Training of the Proposed Relevance Vector Regression (RVR)-Based SOH Estimation Model}
\label{alg:RVR}
\end{algorithm*}

With the optimal set of features found, the next step is to build a module-level SOH estimation model. This section corresponds to Step 4 of Fig. \ref{fig:Algorithm} where the proposed algorithm trains a sparse probabilistic model and estimates module-level SOH as distributions with point estimates and three-sigma credible intervals. 

Relevance vector regression (RVR) is a sparse regression technique and the Bayesian counterpart to non-Bayesian support vector regression \cite{Tipping01}. RVR is chosen in this work over other regression techniques, because it has the following advantages. First, trained models can be updated based on future data using sequential Bayesian learning \cite{Tipping01}. Second, when making predictions, instead of seeking to have accurately trained model parameters, the RVR integrates over all these unknown parameters, which improves predictive performance and increases robustness \cite{Tipping01}. Third, the RVR automatically prefers simple but sufficiently accurate models over complicated models to explain the data \cite{MacKay}. Fourth, kernels in RVR do not need to satisfy Mercer's conditions, allowing higher design freedom \cite{Tipping03}.

\subsection{Regression Algorithm} 
This paper uses the original RVR algorithm \cite{Tipping02}, with customizations in the initialization, kernel choice, hyperparameter values, and stopping conditions to fit the needs of SOH estimation problems. The pseudo-code for training the RVR model is given in Algorithm \ref{alg:RVR}.

Assume independent and identically distributed (I.I.D.) data $\left\{ \left( \boldsymbol{x}_i, y_i \right) \right\}_{i=1}^N$ where $\boldsymbol{x} \in \mathbb{R}^{N_f}$ are features, $N_f$ is the number of features, $y \in \mathbb{R}$ is the output (i.e., SOH), $i$ is an index, and $N$ is the number of data. Then, an output $y$ and an input $\boldsymbol{x}$ can be related by:
\begin{equation}
    y = \left( w_0 + \sum_{i=1}^{N} w_i K\left( \boldsymbol{x},\boldsymbol{x}_i \right) \right) + \eta, \label{eqn:RVRmodel}
\end{equation}
where $\boldsymbol{x}_i$ are the input data, $w_i$ and $w_0$ are weights and offset to be learned, $\eta$ is the noise in the process which is assumed to be a white Gaussian noise with an unknown variance $\beta^{-1}$ to be learned, and $K\left(\cdot, \cdot \right)$ is a user-defined kernel \cite{Tipping02}. This paper uses the radial basis function for the kernel, i.e., 
\begin{equation}
    K\left(\boldsymbol{x}, \boldsymbol{x^\prime} \right) = \exp{\left( -\rho \| \boldsymbol{x}- \boldsymbol{x^\prime} \|^2_2\right)}, \label{eqn:kernel}
\end{equation}
where $\| \boldsymbol{x}- \boldsymbol{x^\prime} \|_2$ is the Euclidean norm of $\boldsymbol{x}- \boldsymbol{x^\prime}$ and $\rho$ is a tunable hyperparameter. 

The first step in the RVR is to choose prior distributions \cite{Tipping02}. Priors are required for any unknowns one wants to learn and add regularization effects into the training process \cite{Robert}. The priors for RVR are: 
\begin{eqnarray}
    p\left( \boldsymbol{w}|\boldsymbol{\alpha} \right) &=& \prod_{i=0}^N \mathcal{N}\left( w_i | 0, \alpha_i^{-1} \right), \label{eqn:WeightPrior} \\ 
    p\left(\boldsymbol{\alpha} \right) &=& \prod_{i=0}^N \Gamma\left( \alpha_i | a, b \right), \label{eqn:VariancePrior} \\
    p\left( \beta \right) &=& \Gamma\left(\beta | c,d\right), \label{eqn:NoisePrior}
\end{eqnarray}
where $p(\cdot)$ is the probability distribution function, $\boldsymbol{w} = \begin{bmatrix} w_0 & ... & w_N \end{bmatrix}^T \in \mathbb{R}^{N+1}$ are unknown offset and weights, $\boldsymbol{\alpha} = \begin{bmatrix} \alpha_0 & ... & \alpha_N \end{bmatrix}^T \in \mathbb{R}^{N+1}$ are unknown reciprocals of variances for offset and weights, $\mathcal{N}(\cdot|\cdot, \cdot)$ and $\Gamma(\cdot|\cdot, \cdot)$ are the Gaussian and Gamma distributions respectively, and $a$, $b$, $c$, $d$ are small numbers so that $p\left(\boldsymbol{\alpha} \right)$ and $p\left( \beta \right)$ are flat noninformative priors that do not provide any information to the learning algorithm and, thus, allow the data to speak for themselves \cite{Robert}. Following \cite{Tipping02}, this paper chooses $a\rightarrow 0$, $b\rightarrow 0$, $c\rightarrow 0$, $d\rightarrow 0$. Priors in Equations (\ref{eqn:WeightPrior}) and (\ref{eqn:VariancePrior}) achieve model sparsity, because these two priors together make the unconditional prior $p\left(\boldsymbol{w} \right)$ a student's t distribution \cite{Tipping02}, a type of very sparse prior distributions \cite{Steinke}. 

The next step is to choose a likelihood distribution for the given data \cite{Tipping02}. Based on the assumptions of I.I.D. sample points and white Gaussian noises in the process, the likelihood for RVR is: 
\begin{equation}
    p\left( \boldsymbol{y} | \boldsymbol{w} ,\beta \right) = \mathcal{N}\left( \boldsymbol{y} | \boldsymbol{\Phi} \boldsymbol{w}, \beta^{-1} \boldsymbol{I}\right),
\end{equation}
where $\boldsymbol{y} = \begin{bmatrix} y_1 & ... & y_N \end{bmatrix}^T$ are outputs, $\boldsymbol{\Phi} = \begin{bmatrix} \boldsymbol{\phi} \left(\boldsymbol{x}_1 \right) & ... & \boldsymbol{\phi} \left(\boldsymbol{x}_N \right) \end{bmatrix}^T \in \mathbb{R}^{N \times \left(N+1\right)}$ is the kernel matrix, $\boldsymbol{\phi} \left(\boldsymbol{x}_i\right) = \begin{bmatrix} 1 & K\left(\boldsymbol{x}_i,\boldsymbol{x}_1\right) & ... & K \left(\boldsymbol{x}_i,\boldsymbol{x}_N\right)\end{bmatrix}^T$ is a vector that contains kernel functions, and $\boldsymbol{I}$ is the identity matrix.

With priors and likelihood defined, the most fundamental step in Bayesian learning is to compute and update the posterior distribution. Posterior can be computed analytically if a pair of conjugate prior and likelihood is used or approximated using various established methods \cite{Robert}. For the RVR \cite{Tipping02}, the posterior is: 
\begin{equation}
    p\left( \boldsymbol{w}, \boldsymbol{\alpha}, \beta | \boldsymbol{y} \right) = p\left( \boldsymbol{w} | \boldsymbol{y},\boldsymbol{\alpha},\beta \right) p \left( \boldsymbol{\alpha},\beta | \boldsymbol{y} \right), 
\end{equation}
where $p\left( \boldsymbol{w} | \boldsymbol{y},\boldsymbol{\alpha},\beta \right)$ can be found analytically because of the conjugate pair chosen, i.e., 
\begin{equation}
    p\left( \boldsymbol{w} | \boldsymbol{y},\boldsymbol{\alpha},\beta \right) = \mathcal{N} \left( \boldsymbol{w} | \boldsymbol{\mu}, \boldsymbol{\Sigma} \right), 
\end{equation}
where one defines a diagonal matrix $\boldsymbol{A} = \text{diag}\left( \boldsymbol{\alpha} \right)$ and 
\begin{eqnarray}
    \boldsymbol{\Sigma} &=& \left( \beta \boldsymbol{\Phi}^T \boldsymbol{\Phi} + \boldsymbol{A} \right)^{-1}, \label{eqn:SigmaMat} \\
    \boldsymbol{\mu} &=& \beta \boldsymbol{\Sigma} \boldsymbol{\Phi}^T \boldsymbol{y}. \label{eqn:muVec}
\end{eqnarray} 
$p \left( \boldsymbol{\alpha},\beta | \boldsymbol{y} \right)$ cannot be directly evaluated, but can be approximated. When performing prediction, the effects of $p \left( \boldsymbol{\alpha},\beta | \boldsymbol{y} \right)$ can be approximated well by $\delta \left( \boldsymbol{\alpha}_{\text{MP}},\beta_{\text{MP}} \right)$ where $\delta(\cdot,\cdot)$ is the delta function and $\boldsymbol{\alpha}_{\text{MP}}$ and $\beta_{\text{MP}}$ can be found by solving the Type-II maximum likelihood \cite{Tipping02}. Specifically, because $p\left(\boldsymbol{\alpha}\right)$ and $p\left(\beta\right)$ are noninformative priors by construction, the final formulation for the Type-II maximum likelihood problem is: 
\begin{eqnarray}
\boldsymbol{\alpha}_{\text{MP}}, \beta_{\text{MP}} &=& \operatorname*{argmax}_{\boldsymbol{\alpha},\beta} p\left(\boldsymbol{y}|\boldsymbol{\alpha},\beta \right), \label{eqn:maximum_likelihood_opt}\\
p\left(\boldsymbol{y}|\boldsymbol{\alpha},\beta\right) &=& \mathcal{N}\left( \boldsymbol{y} | \boldsymbol{0}, \beta^{-1} \boldsymbol{I} + \boldsymbol{\Phi}\boldsymbol{A}^{-1} \boldsymbol{\Phi}^T \right),
\end{eqnarray}
where $\boldsymbol{0}$ is a vector with all components equal to 0. Thus, the iterative update laws for $\boldsymbol{\alpha}$ and $\beta^{-1}$ are: 
\begin{eqnarray}
    \alpha_i &=& \frac{\gamma_i}{\mu_i^2}, \quad \gamma_i = 1-\alpha_i^{\text{old}} \Sigma_{ii}, \label{eqn:alpha_update}\\
    \beta^{-1}  &=& \frac{\left\| \boldsymbol{y}- \boldsymbol{\Phi} \boldsymbol{\mu} \right\|^2_2}{N - \sum_i \gamma_i}, \label{eqn:beta_update}
\end{eqnarray}
where $\Sigma_{ii}$ and $\mu_i$ are the $i$-th diagonal component of matrix $\boldsymbol{\Sigma}$ and the $i$-th component of vector $\boldsymbol{\mu}$ in Equations (\ref{eqn:SigmaMat}) and (\ref{eqn:muVec}) respectively, and $\sum$ in the denominator of Equation (\ref{eqn:beta_update}) is the summation operator. 

Note that, as iterations progress, most of the $\alpha_i$ approach $\infty$, which makes their corresponding weights $w_i$ have distributions with zero mean and zero variance. These $w_i$ and the corresponding components of $\boldsymbol{\Sigma}$, $\boldsymbol{\mu}$, $\boldsymbol{\Phi}$, $\boldsymbol{\alpha}$, and $\boldsymbol{\alpha}^{\text{old}}$ need to be deleted from iterations \cite{Tipping02}. The $\boldsymbol{x_i}$ corresponding to the remaining $w_i$ with nonzero variances are the relevance vectors. This process is called the automatic relevance determination \cite{Neal}. 

When predicting outputs, the key strength of Bayesian learning is the integration over unknown parameters. In RVR \cite{Tipping02}, given a new input $\boldsymbol{x}$, the distribution of estimated output $y$ is: 
\begin{equation}
    p\left( y | \boldsymbol{y},\boldsymbol{\alpha}_{\text{MP}},\beta_{\text{MP}} \right) = \int p\left(y | \boldsymbol{w},\beta_{\text{MP}} \right) p\left(\boldsymbol{w} | \boldsymbol{y}, \boldsymbol{\alpha}_{\text{MP}},\beta_{\text{MP}} \right) d\boldsymbol{w}. 
\end{equation}
This integral has the analytical solution: 
\begin{eqnarray}
    p\left( y | \boldsymbol{y},\boldsymbol{\alpha}_{\text{MP}},\beta_{\text{MP}} \right) &=& \mathcal{N} \left( y | t,\sigma^2 \right), \\
    t &=& \boldsymbol{\mu}^T \boldsymbol{\tilde{\phi}} \left(\boldsymbol{x} \right), \label{eqn:mean_estimate}\\
    \sigma^2 &=& \beta^{-1}_{\text{MP}} + \boldsymbol{\tilde{\phi}}\left(\boldsymbol{x} \right)^T \boldsymbol{\Sigma} \boldsymbol{\tilde{\phi}} \left(\boldsymbol{x} \right), \label{eqn:var_estimate}
\end{eqnarray}
where $\boldsymbol{\tilde{\phi}} \left(\boldsymbol{x} \right) = \begin{bmatrix} 1 & K\left( \boldsymbol{x}, \textbf{rv}_1 \right) & ... & K\left( \boldsymbol{x}, \textbf{rv}_{N_{\text{rv}}}\right) \end{bmatrix}^T$ if the offset is used and $\boldsymbol{\tilde{\phi}}\left(\boldsymbol{x} \right) = \begin{bmatrix} K\left( \boldsymbol{x}, \textbf{rv}_1 \right) & ... & K\left( \boldsymbol{x}, \textbf{rv}_{N_{\text{rv}}}\right) \end{bmatrix}^T$ otherwise, $\textbf{rv}_j,j=1,...,N_{\text{rv}}$ are the relevance vectors found, $N_{\text{rv}}$ is the total number of relevance vectors, and the pruned $\boldsymbol{\Sigma}$ and $\boldsymbol{\mu}$ only contain components corresponding to these relevance vectors after training. With estimate distributions found, this paper uses Equation (\ref{eqn:mean_estimate}) as the point estimate and extracts the three-sigma credible interval from Equation (\ref{eqn:var_estimate}), i.e., $\left(t-3\sigma,t+3\sigma\right)$.

Five customizations are made in this study to implement the RVR proposed by \cite{Tipping02}. First, following the implementations in sklearn-RVM \cite{Tipping03}, this paper initializes all the $\alpha_i$ to be $1/\left(N+1\right)^2$ and $\beta^{-1}$ to be $\left(0.1 \cdot \text{std}\left(\{y_i\}_{i=1}^N\right)\right)^{2}$. Second, during the iterative process, this paper performs the aforementioned deletion when $\alpha_i \geq \alpha_{\text{th}}$ where $\alpha_{\text{th}}$ is a very large number (e.g., $10^9$). Third, this paper stops the iterative process when either the maximum number of iterations $N_{\text{iter}}$ is reached or $\left\| \boldsymbol{\alpha} - \boldsymbol{\alpha}^{\text{old}}\right\|_\infty \leq \text{tol}$ where $\| \cdot \|_\infty$ is the $l_\infty$-norm and $\text{tol}$ is a tunable tolerance. Fourth, to help $\boldsymbol{\Sigma}$ and $\boldsymbol{\mu}$ stay well-conditioned during iterations and make Optimization (\ref{eqn:maximum_likelihood_opt}) converge faster, the data $\left\{ \left( \boldsymbol{x}_i, y_i \right) \right\}_{i=1}^N$ are standardized using Equation (\ref{eqn:standardize}) before running Algorithm \ref{alg:RVR}. Fifth, to prevent $\alpha_i$ in Equation (\ref{eqn:alpha_update}) from becoming very small in practice and slow down the convergence, based on the sklearn-RVM \cite{Tipping03}, this paper modifies the iterative update law in Equation (\ref{eqn:alpha_update}) as:  
\begin{equation}
    \alpha_i = \frac{\max \{\gamma_i, \varepsilon \}}{\mu_i^2}, \quad \gamma_i = 1-\alpha_i^{\text{old}} \Sigma_{ii} \label{eqn:alpha_update_new}
\end{equation}
where $\varepsilon$ is a very small number (e.g., $10^{-8}$).

%%%%%%%%%%%%%%%%%%%%%%%%%%%%%%%%%%%%%%%%%%%%%%%%%%%%%%%%%%%%%%%%%%%%%%%%%%%%%%%%
\section{BATTERY DATASETS}
\label{Dataset}

The proposed method in Fig. \ref{fig:Algorithm} will be demonstrated and validated using experimental datasets. First, the public Oxford lithium cobalt oxide (LCO) cell dataset \cite{Birkl} (downloading website: \cite{OxfordDataset}) is used to illustrate related concepts and interpret results from the proposed method. The detailed description can be found in \cite{Birkl}. Then, a proprietary lithium nickel-manganese-cobalt oxide 622 (NMC622) module dataset is used to evaluate the performance of the proposed method for module-level SOH estimation under cell-to-cell variations. The existence of cell-to-cell variation within modules and their impacts on IC curves are illustrated in Fig. \ref{fig:IC_Curve_C2Cvar}. The entire battery pack containing 96 modules in series is cycled under the charging profiles consisting of two consecutive constant-current regimes with 86A and 64.5A pack-level currents, respectively, and one subsequent constant-voltage regime. Only the first constant-current segment is used for this study, as this segment contains all the IC and DV features for this dataset. The actual data used by the proposed method have a 0.033Hz sampling rate, i.e., 1 measurement per 30 seconds. When the entire pack is charged, passive balancing is used for 96 serially connected modules. The balancing current is very small compared to the module-level current, thus its impact on IC and DV curves is negligible, and the issue discussed in \cite{Tang2} is avoided. Within each module, 3 parallelly connected cells have self-balancing effects and do not involve active or passive balancing.

\begin{table}[H]
\centering
\caption{Key Attributes of Battery Datasets}
\label{table:Datasets}
\begin{tabular}{|c|l|} 
\hline
\textbf{Type} & \multicolumn{1}{c|}{\textbf{Attributes}} \\ 
\hline\hline

Cell & \begin{tabular}[c]{@{}l@{}}\textbf{Chemistry:} LCO\\ \textbf{Cell Nominal Capacity:} 0.74Ah\\ \textbf{No. of Cells:} 8 \\ \textbf{No. of Features:} 24\\ \textbf{No. of Cycles:} 398\\ \textbf{Cell-Level SOH Range:} 100\% - 88\%, \\ where all the IC and DV features exist \end{tabular} \\ 
\hline
Module & \begin{tabular}[c]{@{}l@{}}\textbf{Chemistry:} NMC622 \\ \textbf{Module Nominal Capacity:} 208Ah\\ \textbf{Module Configuration:} 3 cells in parallel\\ \textbf{No. of Modules:} 96 \\ \textbf{No. of Features:} 6\\ \textbf{No. of Cycles:} 81216 \\ \textbf{Module-Level SOH Range:} 100\% - 86\% \end{tabular} \\ 
\hline
\end{tabular}
\end{table}

\begin{figure}[H]
    \centering
    \begin{subfigure}[h]{0.48\textwidth}
        \centering
        \includegraphics[width=\textwidth,trim=20 75 60 25,clip]{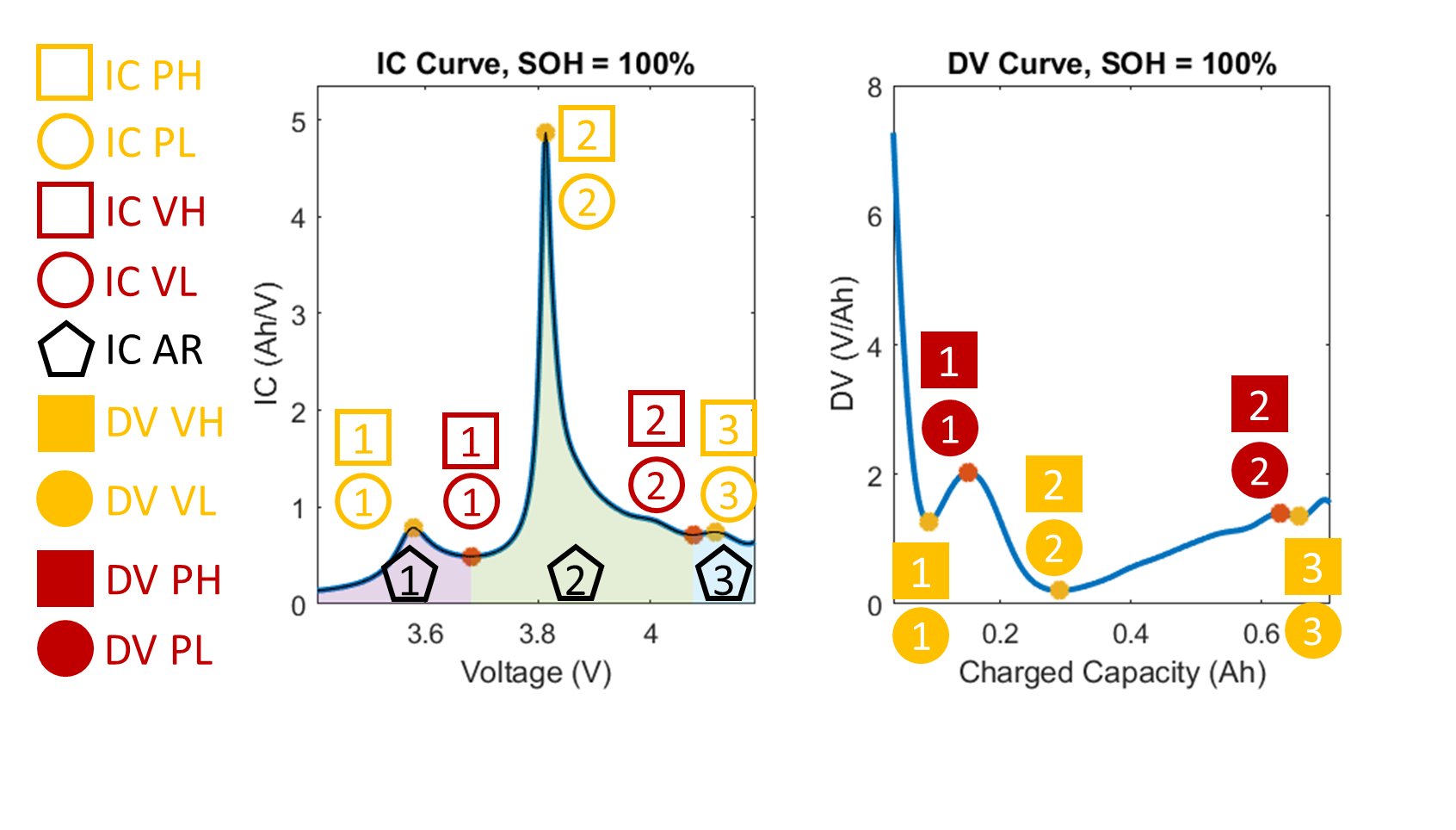}
        \caption{Experimental LCO Cell}
        \label{fig:LCO_Example_IC}
    \end{subfigure} \hfill
    \begin{subfigure}[h]{0.48\textwidth}
        \centering
        \includegraphics[width=\textwidth,trim=0.1 2 60 5,clip]{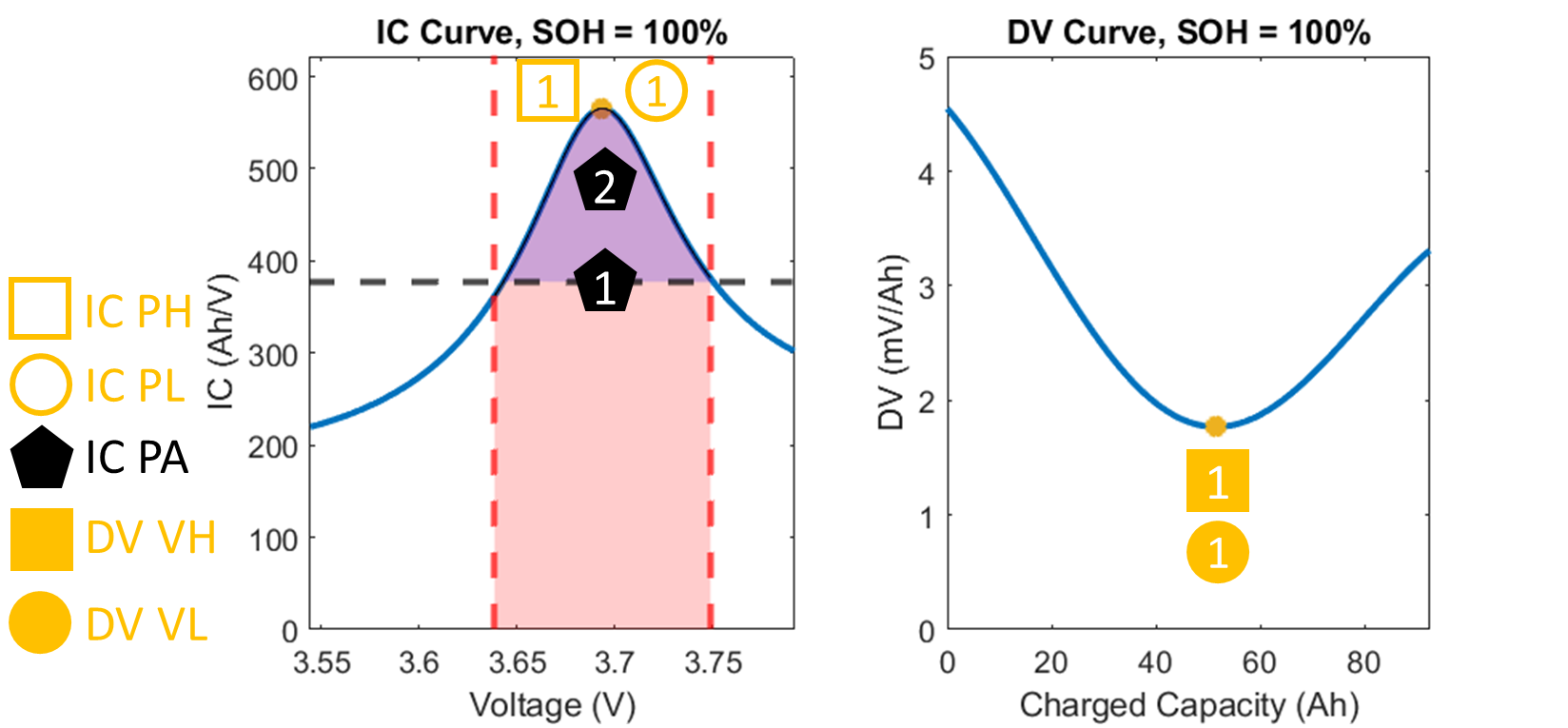}
        \caption{Experimental NMC622 Module}
        \label{fig:NMC622_Example_IC}
    \end{subfigure} \hfill
\caption{Example IC and DV Curves and Related Features for Two Datasets}
\label{fig:Dataset_Features}
\end{figure}

\begin{figure*}[ht]
    \centering
    \begin{subfigure}[h]{0.9\textwidth}
        \centering
        \includegraphics[width=\textwidth,trim=5 5 5 5,clip]{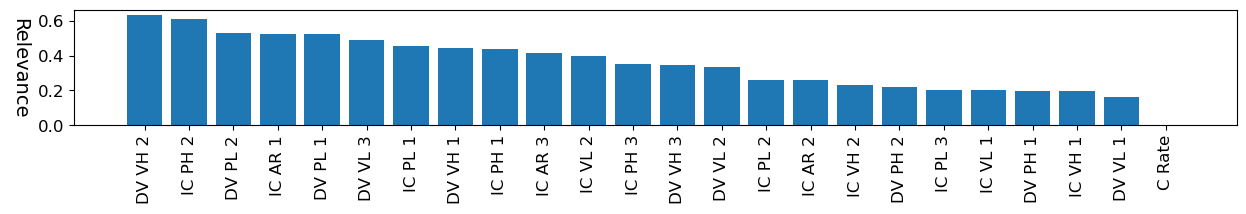}
        \caption{Feature Relevance}
        \label{fig:LCO_Relevance}
    \end{subfigure} \hfill

    \begin{subfigure}[h]{0.45\textwidth}
        \centering
        \includegraphics[width=\textwidth,trim=5 5 4 5,clip]{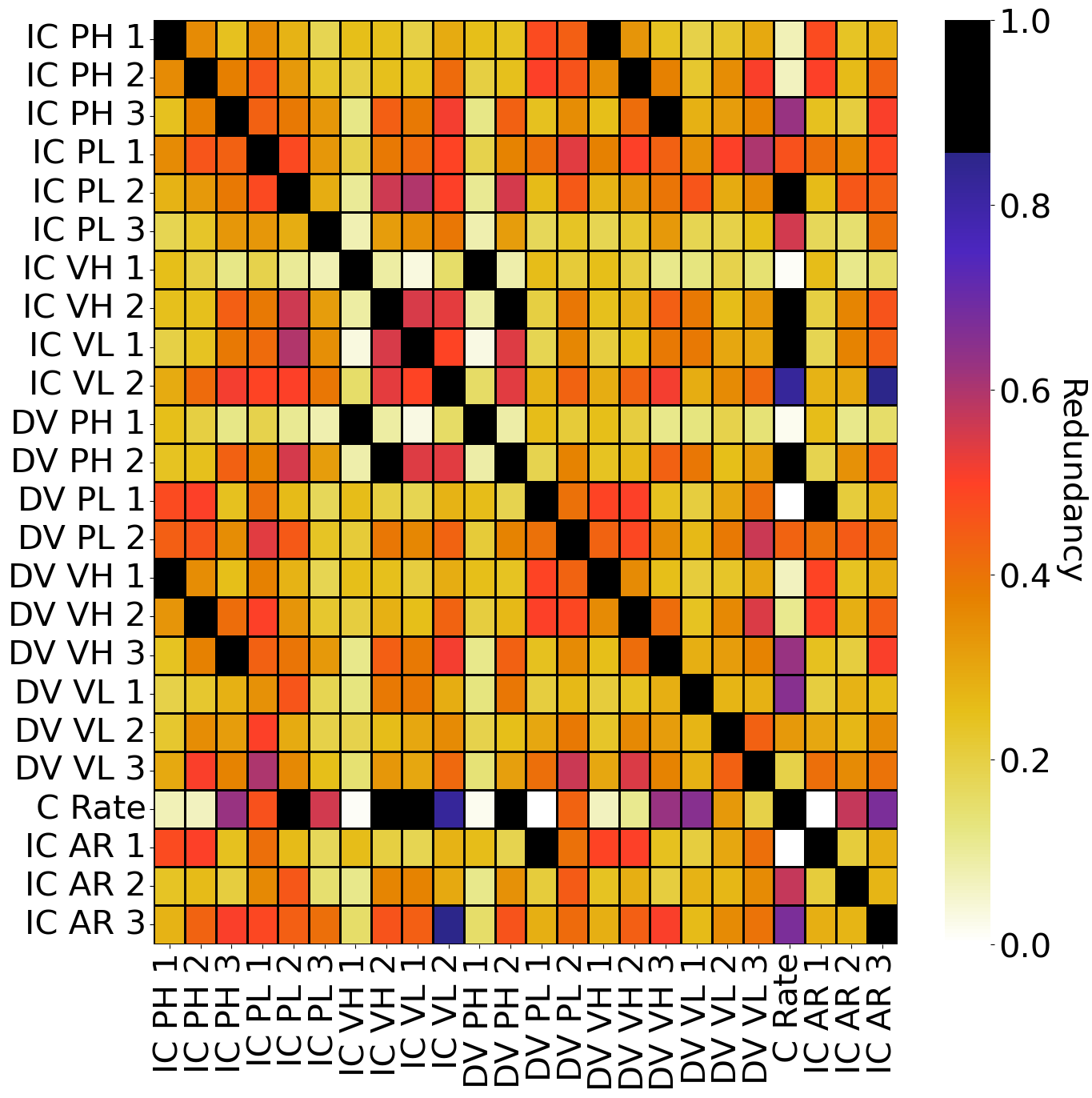}
        \caption{Feature Redundancy}
        \label{fig:LCO_Redundancy}
    \end{subfigure}
    \begin{subfigure}[h]{0.45\textwidth}
        \centering
        \includegraphics[width=\textwidth,trim=5 5 4 5,clip]{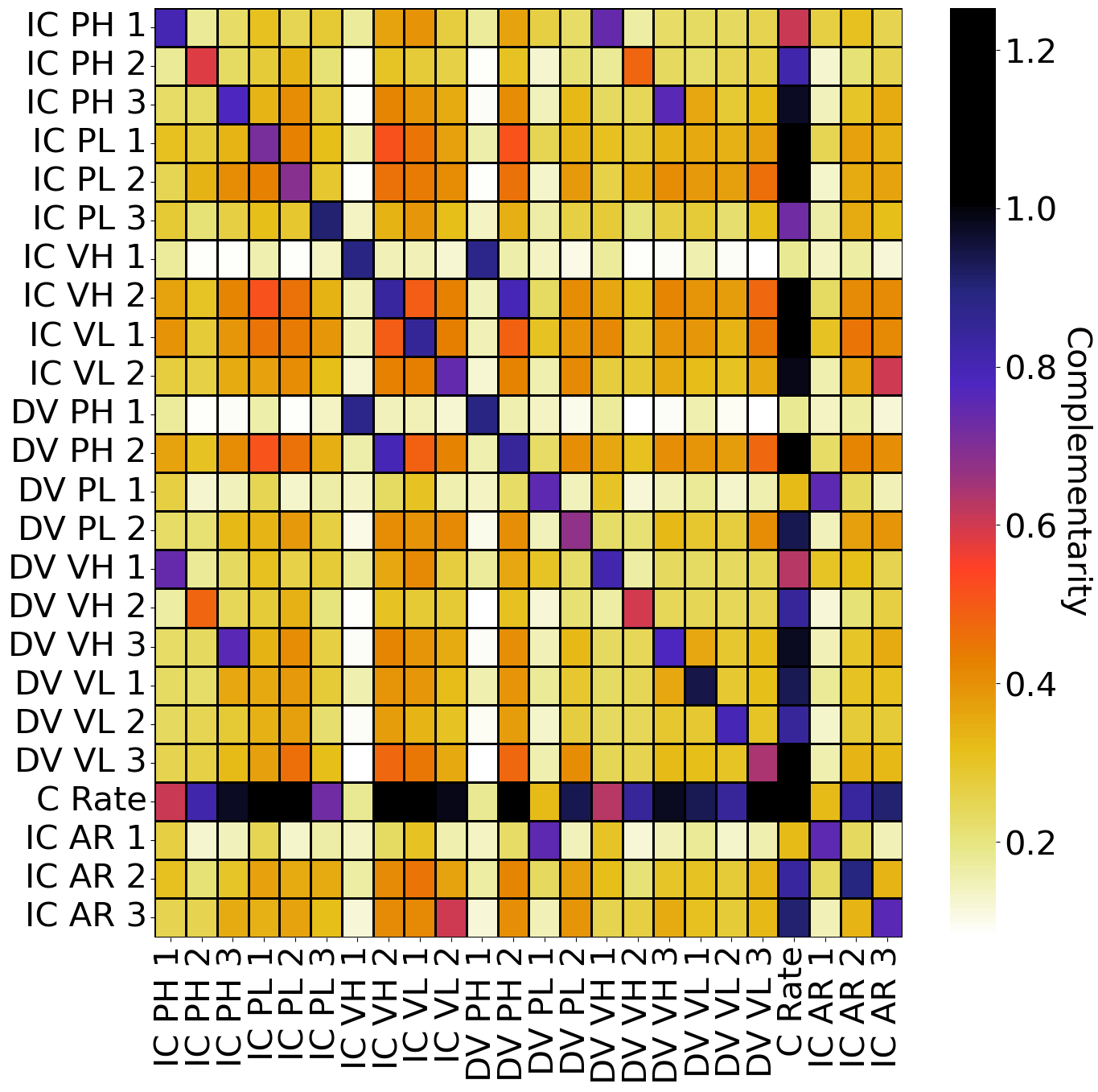}
        \caption{Feature Complementarity}
        \label{fig:LCO_Complementarity}
    \end{subfigure}
\caption{Feature Relevance, Redundancy, and Complementarity for Cell-Level SOH Estimation under Different C Rates}
\label{fig:LCO_FS_results}
\end{figure*}

Table \ref{table:Datasets} summarizes the key attributes of these two datasets, while Fig. \ref{fig:Dataset_Features} shows their example IC and DV curves and related features. Note that, for ease of discussion, the acronyms in Table \ref{table:Features} and indices labeled in Fig. \ref{fig:Dataset_Features} will be used together to refer to different features. For example, the left IC peak height in Fig. \ref{fig:LCO_Example_IC} is denoted as IC PH 1.
%%%%%%%%%%%%%%%%%%%%%%%%%%%%%%%%%%%%%%%%%%%%%%%%%%%%%%%%%%%%%%%%%%%%%%%%%%%%%%%%
\section{INTERPRETATION OF THE PROPOSED METHOD USING CELL DATA}
\label{Cell_Result}
The cell-level SOH estimation is discussed first, because the cell dataset contains a lot more features and charging conditions than the proprietary module dataset, making the cell dataset more suitable to demonstrate the full capability of the proposed feature selection algorithm. Note that the proposed method can be applied to both cell-level and module-level SOH estimations. Nonetheless, as to be seen in Sections \ref{Cell_Result} and \ref{Results}, the selected optimal sets of features and resulting RVR models are different for cell-level and module-level SOH estimations, as the situations and conditions change.

The Oxford cell dataset is split into 80\% training and 20\% testing sets. The training data are used for performing the feature selection and building the RVR-based SOH estimation model, while testing data are used to evaluate the performance. For running the proposed method, as discussed in Sections \ref{CMI_Est} and \ref{RVR}, SOH values and all the features are standardized based on the corresponding means and standard deviations of the training set using Equation (\ref{eqn:standardize}). Then, for result discussions, all these standardized quantities are converted back for physical interpretations.

Fig. \ref{fig:LCO_FS_results} shows the relevance, redundancy, and complementarity among all the features. One can observe from Fig. \ref{fig:LCO_Relevance} that: (1) DV VH 2 has the highest relevance to SOH and it will be the first feature selected by the algorithm when there is no pre-selected feature; (2) C rate has no relevance to SOH, because physically a battery can be at any SOH when charged at a given C rate. 

The black color in Fig. \ref{fig:LCO_Redundancy} represents redundancy values above the threshold $\tilde{I}_{\text{th}}$. The darker the color is, the higher the redundancy is. From Fig. \ref{fig:LCO_Redundancy}, one can observe the following. First, IC peak/valley heights and DV valley/peak heights are completely redundant, because $\text{IC} = 1/\text{DV}$. Second, DV PL 1 and IC AR 1 are completely redundant. From the definition, $\text{IC AR 1} = \int_{V_{\text{min}}}^{V_{\text{IC valley 1}}} \frac{dQ_c}{dV} dV = Q_{c,\text{IC valley 1}} - 0 = \text{DV PL 1}$, where $Q_c$ is the charged capacity, $V$ is the voltage. Third, the C Rate is completely redundant to IC PL 2, IC VH 2, IC VL 1, and DV PH 2. Note that this observation is only applicable to this dataset.

\begin{table*}
\centering
\caption{Feature Selection Results for Cell-Level SOH Estimation of LCO Cells under Different C Rates}
\label{Table:FS_Results_LCO}
\begin{tabular}{|l|llllllll|} 
\hline
\textbf{Ranked Selected Feature Set, $\mathcal{S}$} & \{ & DV VH 2, & C Rate, & DV VL 3, & DV VH 1, & DV PL 2, & IC PL 1, &  \\
 &  & IC AR 3, & DV PL 1, & IC VL 2, & IC PH 3, & DV VL 2, & IC AR 2, &  \\
 &  & IC PL 3, & DV VL 1, & DV PH 1 &  &  &  & \} \\ 
\hline \hline
\textbf{Unranked Removed Feature Set, $\mathcal{R}$} & \{ & IC PH 1, & IC PH 2, & IC PL 2, & IC VH 1, & IC VH 2, & IC VL 1, & \\
 & & DV PH 2, & DV VH 3, & IC AR 1  &  & &  & \} \\
\hline
\end{tabular}
\end{table*}

Based on Fig. \ref{fig:LCO_Complementarity}, complementarity between the C rate and most of the IC and DV features is high, because the values of most features have C rate dependency. Thus, in the case where these features are selected, the C rate will be selected because of the complementarity in Equation (\ref{eqn:FS_obj}), unless the C rate is completely redundant to the associated feature. Note that, from Fig. \ref{fig:LCO_Complementarity}, completely redundant features also have high CMI, but they will not be selected because of the feature removal process in Algorithm \ref{alg:infoFS}.

\begin{remark}\label{remark_on_CMI_norm}
    (On Normalization of CMI) The denominator of Normalization (\ref{eqn:CMI_norm}) is $\min \left(I\left(F;F\right),I\left(G;G\right)\right)$, instead of $\min \left(I\left(F;F|H\right),I\left(G;G|H\right)\right)$, because the normalized value using $\min \left(I\left(F;F|H\right),I\left(G;G|H\right)\right)$ is less interpretable in some situations of this case study. For example, consider the normalized CMI $\tilde{I}\left(\text{IC PH 1}; \text{DV VH 1} | \text{SOH} \right)$. Given $\tilde{I}\left(\text{IC PH 1}; \text{SOH}\right) = 0.4364 > 0$, logically the CMI should be a lot less than 1, because the uncertainty of IC PH 1 has been already reduced by knowing SOH. Definition (\ref{eqn:CMI_norm}) gives $\text{CMI} =  0.7415$, but normalization with respect to $\min \left(I\left(F;F|H\right),I\left(G;G|H\right)\right)$ gives $\text{CMI} = 0.9242 \approx 1$.
\end{remark}

\begin{remark}\label{remark_on_values_of_CMI_norm}
    (On Values of Normalized CMI) Note that, because of the estimation inaccuracy of the used CMI estimator, the normalized CMI could occasionally become a little above 1, as shown in Fig. \ref{fig:LCO_Complementarity}. 
\end{remark}

With relevance, redundancy, and complementarity obtained above, Algorithm \ref{alg:infoFS} outputs the ranked selected feature set and unranked removed feature set, as summarized in Table \ref{Table:FS_Results_LCO}. Note that the DV VH 2 (equivalently, IC PH 2) and C rate are the top two ranked features, which matches the physical knowledge from \cite{Zhou1}. 

\begin{figure}[H]
    \centering
    \includegraphics[width=0.46\textwidth,trim=5 5 5 5,clip]{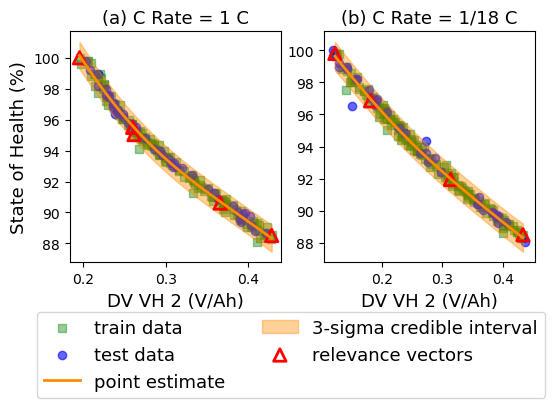}
    \caption{Cell-Level SOH Estimation Model Using Two Features}
    \label{fig:C_rate_LCO_model}
\end{figure}

These top two ranked features are used to build a cell-level SOH estimation model. Based on Table \ref{table:LCO_C_Rate_SOH_Results}, the proposed model has a low computational complexity (small number of relevance vectors, $\textbf{rv}_j$) and high estimation accuracy (low test root-mean-square error, RMSE). Fig. \ref{fig:C_rate_LCO_model} shows the trained model under different C rates and the three-sigma (99.7\%) credible intervals for different SOHs. Table \ref{table:LCO_C_Rate_SOH_Results} also summarizes the average three-sigma values.

\begin{table}[H]
\centering
\caption{Cell-Level SOH Estimation Results}
\label{table:LCO_C_Rate_SOH_Results}
\begin{tabular}{|c|c|c|c|} 
\hline
\begin{tabular}[c]{@{}c@{}}\textbf{Train}\\\textbf{RMSE}\end{tabular} & \begin{tabular}[c]{@{}c@{}}\textbf{Number}\\\textbf{of} $\textbf{rv}_j$ \end{tabular} & \begin{tabular}[c]{@{}c@{}}\textbf{Average}\\\textbf{Three-Sigma Values}\end{tabular} & \begin{tabular}[c]{@{}c@{}}\textbf{Test }\\\textbf{RMSE}\end{tabular} \\ 
\hline \hline
0.27\% SOH & 9 & 0.84\% SOH & 0.33\% SOH \\
\hline
\end{tabular}
\end{table}

\begin{figure*}[ht]
\centering
\begin{subfigure}{0.3\textwidth}
\includegraphics[width=\textwidth,trim=4 5 4 5,clip]{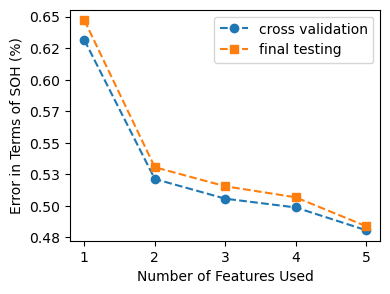}
\caption{Estimation RMSE}
\label{fig:NMC622_RMSE}
\end{subfigure}\hfill
\begin{subfigure}{.3\textwidth}
\includegraphics[width=\textwidth,trim=4 5 4 5,clip]{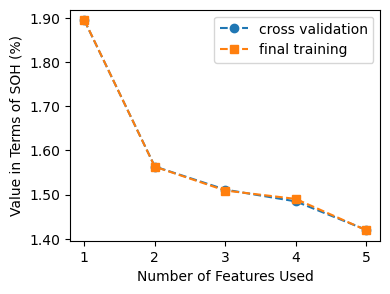}
\caption{Average Three-Sigma Values}
\label{fig:NMC622_ThreeSigma}
\end{subfigure}\hfill
\begin{subfigure}{.3\textwidth}
\includegraphics[width=\textwidth,trim=4 5 4 5,clip]{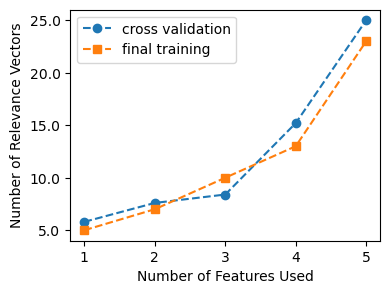}
\caption{Number of Relevance Vectors}
\label{fig:NMC622_N}
\end{subfigure}

\caption{Module-Level SOH Estimation Performance under Different Number of Features Used}
\label{fig:NMC622_SOH_Results}
\end{figure*}
%%%%%%%%%%%%%%%%%%%%%%%%%%%%%%%%%%%%%%%%%%%%%%%%%%%%%%%%%%%%%%%%%%%%%%%%%%%%%%%%
\section{MODULE-LEVEL SOH ESTIMATION RESULTS UNDER CELL-TO-CELL VARIATIONS}
\label{Results}

This section presents the results of module-level SOH estimation in the presence of cell-to-cell variations, using the experimental NMC622 module data to demonstrate real-world performance. Furthermore, this section investigates the onboard computational footprint of the proposed method.

\subsection{Module-Level SOH Estimation Performance} 
\label{SOHm_Reuslts} 
4060 sample points are randomly selected from the large dataset for both feature selection and RVR model development, while the remaining 77156 sample points are used for testing the performance. Results from the two-sample Kolmogorov-Smirnov test \cite{Hodges}, Anderson-Darling test \cite{Scholz}, and Cramér-von Mises criterion \cite{Anderson} all indicate that the randomly selected 4060 sample points have the same distribution as the original dataset. To apply the proposed method, due to reasons discussed in Sections \ref{CMI_Est} and \ref{RVR}, SOH values and all the features of the entire dataset are standardized based on the corresponding means and standard deviations of the training set using Equation (\ref{eqn:standardize}). Then, all these standardized quantities are converted back, in result discussions, for physical interpretations.

\begin{table}[H]
\centering
\caption{Feature Selection Results for Module-Level SOH Estimation in the Presence of Cell-to-Cell Variations}
\label{table:FS_Results_NMC622}
\begin{tabular}{|c||c|} 
\hline
\textbf{Ranked Selected Feature Set, $\mathcal{S}$} & \begin{tabular}[c]{@{}c@{}} IC PA 1 \\ IC PA 2 \\ IC PH 1 \\ DV VL 1 \\ IC PL 1 \end{tabular} \\ 
\hline
\textbf{Unranked Removed Feature Set, $\mathcal{R}$} & DV VH 1 \\
\hline
\end{tabular}
\end{table}

The proposed Algorithm \ref{alg:infoFS} finds the optimal set of features for module-level SOH estimation in the presence of cell-to-cell variations. Table \ref{table:FS_Results_NMC622} summarizes the module-level feature selection results. IC PA 1 is the top-ranked feature, which is not only related to SOH but also the most insensitive to cell-to-cell variations. Other features are selected, according to the trade-off they provide among relevance, redundancy, and complementarity, instead of how insensitive they are to cell-to-cell variations. Referring back to the issue shown in Fig. \ref{fig:IC_Curve_C2Cvar}, even though IC PH 1 is significantly distorted by cell-to-cell variations, the proposed method finds that IC PH 1 provides the best trade-off among relevance, redundancy, and complementarity, once IC PA 1 and IC PA 2 are selected. Thus, instead of using IC PH 1 as a sole feature for SOH estimation, the proposed algorithm recommends IC PH 1 to be used together with IC PA 1 and IC PA 2 to provide optimal estimation performance for this dataset.

\begin{table*}[ht]
\centering
\caption{Module-Level SOH Estimation Results for Modules with Cell-to-Cell Variations under Different Number of Features Used}
\label{table:NMC622_Estimation_Results}
\begin{tabular}{|c|c|c|c|c|c|} 
\hline
\begin{tabular}[c]{@{}c@{}}\textbf{Model}\\\textbf{Name}\end{tabular} & \begin{tabular}[c]{@{}c@{}}\textbf{Number of}\\\textbf{Features}\end{tabular} & \textbf{Selected Features} & \begin{tabular}[c]{@{}c@{}}\textbf{Number of}\\ \textbf{Relevance Vectors} \end{tabular} & \begin{tabular}[c]{@{}c@{}}\textbf{Average}\\\textbf{Three-Sigma Values}\end{tabular} & \begin{tabular}[c]{@{}c@{}}\textbf{Test }\\\textbf{RMSE}\end{tabular} \\ 
\hline \hline
Model 0 & 1 & \begin{tabular}[c]{@{}c@{}} \{IC PH 1\} \\(commonly used in the literature) \end{tabular} & 23 & 3.88\% SOH & 1.29\% SOH \\
\hline 
Model 1 & 1 & \begin{tabular}[c]{@{}c@{}} \{IC PA 1\} \\(selected by the proposed algorithm) \end{tabular} & 5 & 1.90\% SOH & 0.65\% SOH \\
\hline
Model 2 & 2 & \{IC PA 1, IC PA 2\} & 7 & 1.56\% SOH & 0.53\% SOH \\
\hline
Model 3 & 3 & \{IC PA 1, IC PA 2, IC PH 1\} & 10 & 1.51\% SOH & 0.52\% SOH \\
\hline
Model 4 & 4 & \{IC PA 1, IC PA 2, IC PH 1, DV VL 1\} & 13 & 1.49\% SOH & 0.51\% SOH \\
\hline
Model 5 & 5 & \{IC PA 1, IC PA 2, IC PH 1, DV VL 1, IC PL 1\} & 23 & 1.42\% SOH & 0.48\% SOH \\
\hline
\end{tabular}
\end{table*}

With all the available features ranked, the SOH estimation model is trained using Algorithm \ref{alg:RVR}. The proper number of features for the SOH estimation model can be determined by applying five-fold cross-validation to the training data. The average validation RMSE, the average number of relevance vectors, and the average three-sigma values from the five-fold cross-validation are used to evaluate the estimation accuracy, model complexity, and estimation uncertainty, respectively. 

The final SOH estimation model is trained using all the training data, and Table \ref{table:NMC622_Estimation_Results} summarizes the final testing RMSE, number of relevance vectors, and average three-sigma values when different numbers of features are used. Note that IC PH 1 is the feature commonly used in the literature \cite{Weng1,Weng2,Weng3,Zhou1} and, thus, Model 0 in Table \ref{table:NMC622_Estimation_Results} will be used as the baseline for discussion. However, IC PH 1 is only ranked in third place by the feature selection algorithm. Based on Table \ref{table:NMC622_Estimation_Results}, even the SOH estimation model using only one feature selected by the proposed algorithm (i.e., Model 1) can significantly outperform the conventional SOH estimation model (i.e., Model 0). From another perspective, these results also exemplify the significant extent to which cell-to-cell variations can impact SOH estimation. The most robust feature against cell-to-cell variation in this dataset (i.e., IC PA 1) can achieve roughly 1/2 of both the test RMSE and average three-sigma value that a sensitive feature (i.e., IC PH 1) can achieve.

\begin{figure}[H]
    \centering
    \includegraphics[width=0.4\textwidth,trim=5 5 5 5,clip]{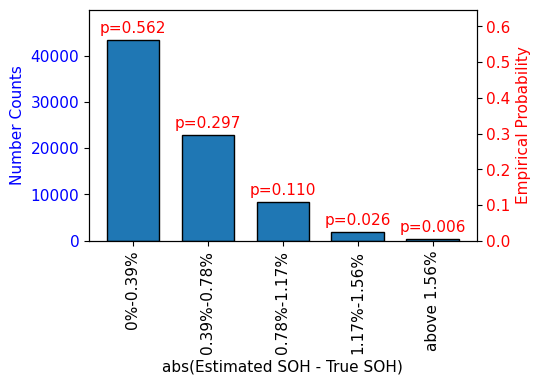}
    \caption{Testing Absolute Error Distribution from Two-Feature SOH Estimation Model}
    \label{fig:Two_Feature_NMC622_Results}
\end{figure}

To reflect the trade-off between estimation performance and model complexity, Fig. \ref{fig:NMC622_SOH_Results} shows how the RMSE, number of relevance vectors, and average three-sigma values vary with the number of features used in the SOH estimation model for cross-validation and testing. Fig. \ref{fig:Two_Feature_NMC622_Results} shows the absolute estimation error distribution for all the testing data when only two features are used. Based on Table \ref{table:NMC622_Estimation_Results}, Fig. \ref{fig:NMC622_SOH_Results}, and Fig. \ref{fig:Two_Feature_NMC622_Results}, several important conclusions could be made: 
\begin{itemize}
    \item Leveraging properly selected multiple features indeed improves module-level SOH estimation accuracy and reduces estimation uncertainty. 
    \item Better performance for module-level SOH estimation can be achieved at the expense of higher model complexity. Using only two or three features provides a good trade-off between estimation performance and model complexity. 
    \item Module-level SOH can indeed be estimated with high accuracy and confidence in the presence of cell-to-cell variations. Note that the proposed method does not use any information about cell-to-cell variations explicitly. 
\end{itemize}

\begin{remark}\label{remark_on_when_tp_estimate_SOH}
    For the proposed SOH estimation method, charging ranges do not necessarily have to be full but have to be long enough to include the selected features. Since the SOH estimation, unlike the state of charge estimation, does not need to be performed continuously for all operations, a condition can be included in the algorithm so that SOH estimation is only performed when there are sufficient data points and bypassed when charging ranges are short.
\end{remark}
%%%%%%%%%%%%%%%%%%%%%%%%%%%%%%%%%%%%%%%%%%%%%%%%%%%%%%%%%%%%%%%%%%%%%%%%%%%%%%%%
\subsection{Onboard Implementation of the Proposed Method}
\label{OnboardImplementation} 
The proposed method in Fig. \ref{fig:Algorithm} has both offboard and onboard computations involved. The proposed feature selection algorithm in Step 3 of Fig. \ref{fig:Algorithm} is performed completely offboard to find the optimal set of features for module-level SOH estimation. The onboard computation involves: (1) extracting the values of these selected features from module-level IC and DV curves and (2) estimating SOH according to the RVR model developed in Step 4 of Fig. \ref{fig:Algorithm}. 

The proposed RVR-based SOH estimation model is sparse, namely, a small number of relevance vectors and one offset scalar need to be stored onboard. Only Equations (\ref{eqn:mean_estimate}) and (\ref{eqn:var_estimate}) need to be implemented onboard. These two equations involve low-dimensional matrix multiplications. Take the NMC622 modules as an example. Based on Section \ref{SOHm_Reuslts}, if one uses two features to estimate SOH, seven 2-by-1 relevance vectors and one offset scalar will be stored onboard and used for all 96 modules. Then, for each module, Equations (\ref{eqn:mean_estimate}) and (\ref{eqn:var_estimate}) involve matrix multiplications among four 8-by-1 vectors and one 8-by-8 matrix.

In summary, the onboard computational footprint of the proposed method in Fig. \ref{fig:Algorithm} is small and involves only extracting feature values and performing low-dimensional matrix multiplications. All the computationally intensive optimization and training processes are done offboard. Moreover, SOH monitoring does not need to be performed continuously. Thus, its implementation will not impose special onboard computational requirements. 
%%%%%%%%%%%%%%%%%%%%%%%%%%%%%%%%%%%%%%%%%%%%%%%%%%%%%%%%%%%%%%%%%%%%%%%%%%%%%%%%
\section{CONCLUSIONS}
\label{Conclusion} 
This paper proposes a novel method and demonstrates the feasibility of estimating module-level SOH with high accuracy and confidence in the presence of cell-to-cell variations. First, an information theory-based feature selection algorithm is proposed to find an optimal set of features for SOH estimation under cell-to-cell variations by optimizing feature relevance, redundancy, and complementarity. The optimal feature set found is independent of subsequent learning algorithms. Second, a relevance vector regression (RVR)-based SOH estimation model is proposed. Model sparsity, high estimation accuracy, and good estimation confidence are demonstrated. Applied to a large dataset, the proposed algorithm achieves module-level SOH estimation with 0.53\% RMSE and 1.56\% average three-sigma value when only two features are used. With more ranked features used in the estimation model, the accuracy can be improved to 0.48\% RMSE and 1.42\% average three-sigma value. Thus, compared to the estimation results (1.3\% RMSE and 3.9\% average three-sigma value) obtained by directly applying IC peak-based methods originally developed for the cell level, the proposed method provides significant improvements in estimation performance. As the optimization and training processes are performed offboard, the developed SOH model can be implemented for onboard execution with low memory and computational requirements.  

Results in Sections \ref{Cell_Result} and \ref{Results} indicate that the same proposed method in Fig. \ref{fig:Algorithm} has good generalizability to work well for both cell-level and module-level SOH estimation problems. Different optimal sets of features will be found by the proposed method for cell-level and module-level SOH estimation problems to give good estimation performance.

Future work on the following topics will be pursued to extend and generalize the results reported in this paper. First, making the feature selection algorithm more robust under different operating conditions will be investigated. For example, two experimental datasets used in Sections \ref{Cell_Result} and \ref{Results} verify the proposed method in a situation where all the originally available features will not disappear throughout the entire degradation process. Future research could be conducted to see how the proposed method performs when some originally available features disappear at some degradation status. Second, how to extend the proposed method to different charging ranges, especially those that do not contain features recommended by the feature selection algorithm, can be investigated in the future. Third, the effects of measurement noise, data acquisition systems, and vehicle chronometrics on the proposed method will be investigated. Fourth, different approaches to estimating cell-to-cell variations within the module, given only module-level measurements, will be explored. Fifth, whether the proposed framework could be applied to estimate the internal resistances of cells and modules for battery power-related performance could also be investigated in the future.


%%%%%%%%%%%%%%%%%%%%%%%%%%%%%%%%%%%%%%%%%%%%%%%%%%%%%%%%%%%%%%%%%%%%%%%%%%%%%%%%

%%%%%%%%%%%%%%%%%%%%%%%%%%%%%%%%%%%%%%%%%%%%%%%%%%%%%%%%%%%%%%%%%%%%%%%%%%%%%%%%

%%%%%%%%%%%%%%%%%%%%%%%%%%%%%%%%%%%%%%%%%%%%%%%%%%%%%%%%%%%%%%%%%%%%%%%%%%%%%%%%
%\section*{APPENDIX}

%\section*{ACKNOWLEDGMENT}

%%%%%%%%%%%%%%%%%%%%%%%%%%%%%%%%%%%%%%%%%%%%%%%%%%%%%%%%%%%%%%%%%%%%%%%%%%%%%%%%
\bibliographystyle{ieeeconf.bst}
\bibliography{ieeeconf.bib}
\end{document}